\documentclass[compsoc,conference,a4paper,10pt,times]{IEEEtran}
\IEEEoverridecommandlockouts
\ifCLASSOPTIONcompsoc
  \usepackage[nocompress]{cite}
\else
  \usepackage{cite}
\fi
\ifCLASSINFOpdf
\else
\fi
\usepackage{amsmath}
\usepackage{enumitem}
\usepackage{filecontents}
\usepackage{tikz} 
\usepackage{tcolorbox}
\usepackage{multirow}
\usepackage{algorithm}
\usepackage{algpseudocode}
\usepackage{float}
\usepackage{graphicx} 
\usepackage{booktabs}
\usepackage[table,xcdraw]{xcolor}
\usepackage{adjustbox}
\usepackage{makecell} 
\usepackage{tabularx}  
\usepackage{diagbox} 

\usepackage{pifont}
\usepackage{hyperref}
\usepackage{pdflscape}
\usepackage{adjustbox}
\usepackage{listings}
\usepackage{xcolor}
\usepackage{tabularx}

\usepackage{url}

\usepackage{breakurl}
\usepackage{xspace} 
\usepackage{soul}
\usepackage{cancel}
\usepackage{tikz}
\usetikzlibrary{calc}

\newcommand{\subhead}[1]{\vspace {0.04in}\noindent{\textbf{#1.}}}
\newcommand{\oursolution}{\textsc{SIABench}\xspace}

\newcommand{\darkcircle}[1]{%
    \tikz[baseline=(char.base)]{
        \node[shape=circle,draw,fill=black,text=white,inner sep=0.5pt,minimum size=1em] (char) {\scriptsize \textbf{#1}};
    }%
}
\newcommand*\circledA[1]{\tikz[baseline=(char.base)]{
            \node[shape=circle,draw,inner sep=1pt,fill=black,text=white,scale=0.8] (char) {#1};}}

\begin{document}
%

\title{Before You Hand Over the Wheel: Evaluating LLMs \\ for Security Incident Analysis}

\author{
\IEEEauthorblockN{
Sourov Jajodia\IEEEauthorrefmark{1},
Madeena Sultana\IEEEauthorrefmark{2},
Suryadipta Majumdar\IEEEauthorrefmark{1},
Adrian Taylor\IEEEauthorrefmark{2},
Grant Vandenberghe\IEEEauthorrefmark{2}
}

\IEEEauthorblockA{\IEEEauthorrefmark{1}
Security Research Centre (SRC)\\
Concordia University, Montreal, Canada\\
sourov.jajodia@mail.concordia.ca\\
suryadipta.majumdar@concordia.ca
}

\IEEEauthorblockA{\IEEEauthorrefmark{2}
Defence Research and Development Canada\\
\{Madeena.Sultana, Adrian.Taylor, Grant.Vandenberghe\}@forces.gc.ca
}
}


%


\maketitle

\begin{abstract}

Security incident analysis (SIA) poses a major challenge for security operations centers, which must manage overwhelming alert volumes, large and diverse data sources, complex toolchains, and limited analyst expertise. These difficulties intensify because incidents evolve dynamically and require multi-step, multifaceted reasoning. Although organizations are eager to adopt Large Language Models (LLMs) to support SIA, the absence of rigorous benchmarking creates significant risks for assessing their effectiveness and guiding design decisions. Benchmarking is further complicated by: (i) the lack of an LLM-ready dataset covering a wide spectrum of SIA tasks; (ii) the continual emergence of new tasks reflecting the diversity of analyst responsibilities; and (iii) the rapid release of new LLMs that must be incorporated into evaluations.

In this paper, we address these challenges by introducing \oursolution, an agentic evaluation framework for security incident analysis. First, we construct a first-of-its-kind dataset comprising two major SIA task categories: (i) deep analysis workflows for security incidents (25 scenarios) and (ii) alert-triage tasks (135 scenarios). Second, we implement an agent capable of autonomously performing a broad spectrum of SIA tasks (including network and memory forensics, malware analysis across binary/code/PDF formats, phishing email and kit analysis, log analysis, and false-alert detection). Third, we benchmark 11 major LLMs (spanning both open- and closed-weight models) on these tasks, with extensibility to support emerging models and newly added analysis scenarios.

\end{abstract}


%
\IEEEpeerreviewmaketitle

\section{Introduction}


Security incident analysis (SIA) is a core responsibility of security 
operations centers (SOCs), yet many SOCs struggle to keep pace with 
the sheer volume and diversity of incidents they must handle~\cite{yang2024true,alahmadi202299}. 
Analysts confront heterogeneous alert types~\cite{nobles2022stress,nist}, 
must pursue multiple analysis objectives, and often rely on a patchwork of tools 
spanning several workflow stages~\cite{mughal2019comprehensive}. 
Although Security Orchestration, Automation, and Response (SOAR) platforms~\cite{splunk_soar,ibm_soar,forti_soar}
provide partial automation, they still demand substantial customization 
and manual effort to handle heterogeneous, evolving incidents~\cite{soar-challenges}. 
As a result, vendors~\cite{adopt_1, adopt_2} have begun exploring the 
use of large language models (LLMs) to augment SIA, motivated by 
LLMs’ ability to perform semantic reasoning across varied security contexts.

However, rushing adoption risks overlooking LLM effectiveness in 
SIA tasks and may ultimately degrade security rather than improve 
it (e.g.,~\cite{secllmholm, capodieci2024impact}). In addition, 
there is a lack of systematic studies evaluating LLM capabilities 
for SIA-specific tasks, unlike the growing body of evaluations in 
other security domains (such as vulnerability management~\cite{llm_vulnerability,secllmholm},
code analysis~\cite{code_analysis,hu2024degpt,usenix25-Spracklen}, threat intelligence 
generation~\cite{alam2024ctibench}, and penetration testing~\cite{deng2024pentestgpt}). 
Existing works on incident response~\cite{lin2025ircopilot,soups_kramer} focus primarily on 
task automation, not on large-scale agentic evaluation of LLMs. A detailed 
literature review appears in Sec.~\ref{sec:relatedwork}. Addressing this gap 
and designing such an evaluation framework introduces several key challenges:

\begin{itemize}[leftmargin=*, itemsep=0pt, topsep=0pt]

\item \textbf{Lack of Real-world SIA Datasets.} 
Real-world security incident datasets remain scarce due to their 
proprietary nature and the significant infrastructure and expertise 
required to build them from scratch~\cite{gennari2024considerations}. 
Unlike other security domains, SIA lacks dedicated CTF-style competitions 
(e.g., picoCTF~\cite{picoctf}, CSAW~\cite{csaw}) from which datasets can 
be directly curated. SIA also involves multi-faceted, iterative workflows 
where analysts must correlate interdependent traces, interpret the broader 
incident context, and adapt to evolving evidence~\cite{dfir}. These tasks require 
deep expertise in system internals, malware analysis, and networking~\cite{cisa, nist}, 
along with proficiency in specialized tools (e.g., Wireshark~\cite{ndatinya2015network}, 
Volatility \cite{daghmehchi2022memory}, Oledump \cite{oledumpppp}).



\item \textbf{Lack of Evaluation Best Practice.} 
Currently, there are no established best practices 
on \emph{what to evaluate} and \emph{how to evaluate} for SIA tasks. 
Unlike CTF exercises, which center on a single goal (capturing the flag), 
SIA comprises multiple objectives such as determining who was involved, 
where the incident occurred, and how it unfolded~\cite{cisa,nist}. 
These tasks must be executed in dynamic environments using a wide range of security tools, 
making systematic evaluation substantially more complex. 
Moreover, the rapid release\footnote{\url{https://nhlocal.github.io/AiTimeline}} 
of new models and variants underscores the need for an evaluation framework 
that can accommodate future models and track progress over time.

\end{itemize}
The following example
further depicts these challenges.

\begin{figure}[ht] 
    \centering
         \vspace{-0.1in}
    \includegraphics[width=\columnwidth]{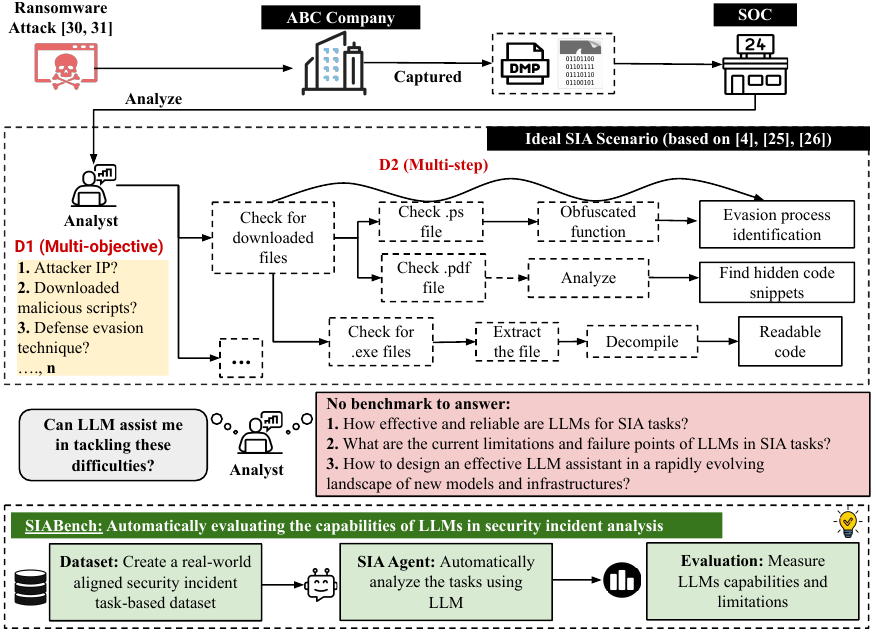} 
    \vspace{-0.3in}
     \caption{Motivating Example.}    
     \vspace{-0.2in}
    \label{fig:Motivating_Example}
\end{figure}

\subhead{Motivating Example}
Fig. \ref{fig:Motivating_Example} depicts a typical security incident scenario in
a SOC environment.
The top of the figure shows a ransomware attack (based on \cite{sans_ransom, flashpoint})
on \texttt{ABC Company}, for which several artifacts (e.g., network logs, and memory dumps) are captured in its
SOC.
The middle of the figure shows an ideal scenario (based on the SIA procedures in \cite{nist, cisa, dfir}), where an analyst 
must follow several complicated steps to analyze this incident. Specifically,
the first difficult task ($D1$) for the
analyst is to address multiple investigative objectives
to reconstruct the attacker’s behavior and provide a complete scope and timeline of the attack.
These include identifying IP addresses associated with scanning patterns (\texttt{Reconnaissance}), finding out the downloaded files (\texttt{Execution through Initial Access}), identifying evasion techniques (\texttt{Defense Evasion}), and more. 
The second difficulty ($D2$) for the analyst
is that this investigative process is not linear and requires 
multiple investigation paths, each providing different pieces of information.
For example, in order to trace \texttt{execution through initial access}, the analyst needs to analyze
\texttt{initial accesses}, like  various types of downloaded files (e.g., \texttt{.pdf} files), 
to find any \texttt{execution} trace like
hidden malicious code snippets; 
along with exploring other investigative paths. 
In addition to these difficulties in SIA, the analyst also suffers
from alert fatigue due to huge number of false positives~\cite{alahmadi202299,yang2024true} 
(not shown in the figure).

From this
desperate need for assistance,
\texttt{ABC Company} (like \cite{adopt_1, adopt_2})
may decide to adopt LLMs to help with SIA.
However,
in the absence of a systematic benchmark,
\texttt{ABC} has limited means to 
answer
to several critical questions
(to help design a solution), such as:
(1) \emph{{how effective and reliable are LLMs for SIA tasks?}},
(2) \emph{{what are the current limitations and failure points of LLMs in SIA tasks?}}, and
(3) \emph{{how to design an effective LLM assistant in a rapidly evolving landscape of new models and infrastructures (e.g., model selection, update strategies, and effective scaffolding)?}}.


To fill this gap, in this paper (as shown at
the bottom of the figure), 
we propose \oursolution, an agentic benchmarking of LLMs for SIA tasks.
Specifically,
we 
develop
LLM-digestible datasets for
SIA tasks, build an agentic framework to automatically 
and continuously perform 
SIA tasks 
and 
systematically evaluate 
the effectiveness of a range of LLMs as a basis for the agent
in SIA.
Thus, this study aims to determine whether LLMs can effectively address the unique challenges of SIA tasks and 
provide insights that help SOCs make informed decisions on adopting LLMs to enhance their security posture.



The key contributions of this paper are as follows:

\begin{itemize} [leftmargin=*, itemsep=0pt, topsep=0pt]

\item \textbf{A Benchmark Dataset~\cite{siabench_link} to Evaluate LLMs on SIA.} 
To the best of our knowledge, we present the first benchmark dataset purpose-built 
for evaluating LLM performance on SIA tasks. The dataset has three key features. 
First, it consists of two major components: (i) an SIA dataset that simulates 
the multi-step investigative workflow of incident analysis
(with 25 incident scenarios and 229 investigative questions), and (ii) an 
alert-triage dataset (with 135 alert scenarios) containing both true and false positives to reflect 
realistic SOC conditions. Second, it spans core SIA domains (network forensics, 
memory forensics, malware analysis, phishing analysis, and log analysis) across 
three difficulty levels ($easy$, $medium$, and $hard$), and is organized around investigative 
questions (e.g., $who$, $what$, $how$~\cite{dfir}) aligned with adversarial tactics from the 
CISA playbook~\cite{cisa}. 
Third, the SIA dataset was carefully curated and pre-processed in consultation with cybersecurity professionals to mitigate bias and enhance real-world fidelity. Additionally, a human expert validation study was conducted to assess the de-biasing process and verify tactic-level assignments.

\item \textbf{An Automated Agent to Perform SIA Tasks.} 
We develop a SIA agent designed to autonomously execute 
SIA tasks for LLM evaluation. The agent provides four key capabilities: 
(i) dynamic access to tools and analysis environments for inspecting incident 
artifacts; (ii) multi-step reasoning across complex, multi-objective incident 
scenarios; (iii) improved long-term planning through selective filtering of 
relevant information from logs and analysis outputs; and (iv) integration 
with 11 major LLMs, with extensibility for additional models and tasks. To assess the capabilities of off-the-shelf LLMs, we intentionally minimized agent scaffolding. An ablation study is included to further justify this design choice.





\item \textbf{Evaluation of Major LLMs for SIA Tasks.} 
We conduct the first systematic and comprehensive evaluation of 11 major LLMs
(four open-weight models
and seven closed-weight models)
on security incident analysis using our datasets and SIA Agent. 
Our results show the continuous progress in LLM performance in SIA tasks
over the last 18 months (where recent models, e.g., Claude-4.5-Sonnet and GPT-5,
significantly outperform the previous models,
especially in more complex SIA tasks) yet leaving a lot of room for improvement. 
We also validate the findings through a practical study on live SIA tasks published after the models’ training cut-off dates.

\end{itemize}

\section{Scope and Assumptions}
In this paper, we evaluate LLMs' performance from the competency perspective of a junior SOC analyst. Our underlying assumption is that if LLMs are to be meaningfully integrated into the automation of SIA tasks, they should first demonstrate the ability to perform at least at the level expected of a junior analyst. This focus reflects the hierarchical nature of SOC, where higher tier works rely heavily on human judgment and oversight, leaving limited scope for full automation. By contrast, the greatest potential for automation lies in lower-tier tasks routinely handled by junior analysts such as alert triaging \cite{radutest, kerstenfield}, assisting senior analyst for deep investigations, etc. To capture these competencies, we construct an alert triage dataset and curate a deep-investigation SIA dataset from established training platforms commonly used to train junior analysts ~\cite{blue_certification, Blue}. This practice is also supported by other benchmarks~\cite{tann2023using,zhang2024cybench,yang2023language} in security, where LLMs' evaluation were conducted on exam questions and CTF competition problems.

Our alert triage task 
data consists of both true and false positives, and we 
evaluate the classification performance of the LLMs using 
our agentic framework. Whereas, our deep investigation 
dataset contains true positives scenarios, where we evaluate
the capability of deep investigation of LLMs in 
real attack scenarios using our framework.
The present scope includes the evaluation of 11 major LLMs, 
with the support for incorporating additional models  
and new security tasks in the future by following the 
procedures outlined in Sec.~\ref{sec:newmodel}.
This work operates under 
three key assumptions: (i) artifacts required for analysis are intact 
and accessible; (ii) LLMs can reliably interact with the command-line 
tools needed to process these artifacts; and (iii) the analysis environment 
has sufficient computational and data resources to support LLM operations.

\section{\oursolution Dataset}
\label{sec:dataset}
Our \oursolution dataset contains two major parts.
\subsection{Part I: Security Incident Analysis (SIA) Tasks}

\begin{figure*}[htp]
    \centering
    \includegraphics[width=\textwidth]{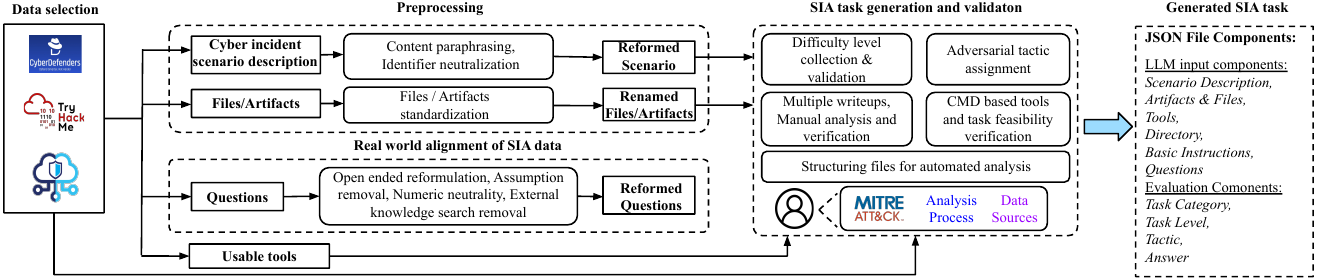}
    \vspace{-0.3in}
    \caption{SIA Dataset Design and Development Process.}
    \label{fig:Question_Formation}
\end{figure*}

Fig. \ref{fig:Question_Formation} outlines the development process of our SIA dataset 
(that includes true security incidents).
The SIA dataset was co-developed with a senior SOC expert (25+ years experience) to ensure real-world alignment, with additional input from two practitioners: an active SOC operator/trainer and another with 6+ years of experience including pen-testing and SANS certifications, who provided insights on workflows, incident reports, and best practices.
Over six months, the team completed four iterative review and validation 
cycles to refine task design and scenario fidelity. We further validate
the dataset with additional 14 security
professionals and researchers via surveys (detailed later).

\subhead{Data Selection}
At first, we carefully curate SIA scenarios from established cyber defense and blue team training platforms \cite{Blue}, including CyberDefenders \cite{CyberDefenders}, Blue Team Labs Online \cite{BTLO}, and TryHackMe \cite{tryhackme}. To maintain the open-source nature of our dataset, we select free scenarios from these three platforms, which are also available via public write-ups (e.g.,~\cite{Panagiotis, forensicskween}). However, a trade-off of using publicly documented scenarios is the potential risk of data contamination such as their inclusion in the training data of large language models, which we address during the pre-processing step (discussed later). Secondly, during scenario selection, we cover the core areas of SIA tasks with varying difficulty levels \cite{dfir, nist}. The selected scenarios encompass a variety of security incidents, including ransomware attacks, web server compromises, phishing-driven malware infections, drive-by download attacks, Linux server compromises, and port scanning attacks. Our dataset can be divided into the following four major categories:

\begin{itemize}[leftmargin=*, itemsep=0pt, topsep=0pt]
    \item \textbf{Network Forensics:} includes tasks requiring the LLMs to analyze malicious network traffic, such as PCAP files.
    \item \textbf{Memory Forensics:} includes tasks requiring the LLMs to
inspect disk images or examine memory dumps.
    \item \textbf{Malware Analysis:} includes tasks, such as malicious file
analysis (PDF, XML, binary, etc.), static code review, and
inspecting ransomware scripts.
    \item \textbf{Miscellaneous:} includes other tasks such as the analysis of
phishing e-mails, log files, and phishing kits.
\end{itemize}
\subhead{Pre-processing} The initial collection of our data includes: description of the scenario, supporting artifacts and files (e.g., .pcap, .mem, .eml files), suggested security tools if available, and key investigation questions for the trainee to conduct the analysis. Given the potential data contamination risk due to the availability of write-ups, we implement a multi-step pre-processing strategy for reducing the likelihood of memorized responses \cite{satvaty2024undesirable, code_analysis}  from the LLMs. 
In addition, we validate our dataset with active SIA tasks that are
published after the models' training cutoff dates (Sec. \ref{subsec:RQ5}). 
We discuss contamination risks in Appendix \ref{sec:discussion}.
\begin{itemize}[leftmargin=*, itemsep=0pt, topsep=0pt]
    \item \textbf{Content Paraphrasing:} We paraphrase each scenario description to minimize the likelihood of exact matches with the LLM's training data. Each scenario description is rephrased using the Gemini 1.5 Flash API\footnote{Gemini 1.5 Flash has been deprecated as of September 2025.}  \cite{Google_AI} using structured prompt (examples in Appendix \ref{sec:prompts}) to ensure scenarios maintain their original integrity while avoiding direct matches with potential training data. 
    

    \item \textbf{Identifier Neutralization:} As certain keywords can trigger LLMs' memorized data \cite{bai2024special}, we remove specific identifiers such as company names and user names, replacing them with generalized terms like \textit{``a company''} or \textit{``an employee''}. Additionally, we reduce unnecessary context by removing lengthy background descriptions, such as detailed attack vector explanations, ensuring the model remains focused on relevant incident scenario information.
    
    \item \subhead{Artifact and File Standardization} 
    Similar to the previous step,
    the original filenames and artifacts for each scenario are renamed using generalized patterns. For example, scenario-specific names like \texttt{lala.pcap} or \texttt{steal.pcap} are changed to neutral names like \texttt{capture.pcap}. This approach, based on findings from prior work \cite{code_analysis} and recommendations from our cybersecurity team, prevents LLMs from triggering any pre-trained knowledge to retrieve memorized answers.

\end{itemize}

\subhead{Real-world Alignment} \label{sec:real_world_alignment}
Our cybersecurity professional observed that the curated investigative questions accompanying the scenarios exhibited several biases and assumptions that might not align with real-world conditions. 
Such biases can influence both human trainees and large language models (LLMs), potentially steering them toward predetermined conclusions \cite{prompt_bias, potential_bias}.  To address this limitation,
we incorporated the expert’s feedback and systematically revised the original questions using four key strategies:


\begin{itemize}[leftmargin=*, itemsep=0pt, topsep=0pt]
    \item \textbf{Open-ended Reformulation:} To encourage evidence-based reasoning and avoid assumptions, we manually rephrase fact-based questions as analytical prompts. For instance, the raw question \textit{``What is the IP responsible for conducting the port scan activity?''} presumes the existence of port scanning. Instead, we ask \textit{``Is there any evidence of port scanning in the network traffic? If so, what is the IP address responsible for the scanning activity?''} This encourages the model to verify suspicious activity before identifying specific indicators.

    \item\textbf{Assumption Removal:} We eliminate specific keyword assumptions (e.g., specific mention of PHP URL, PowerShell Script, Compressed Downloaded File) that could bias model responses to a specific conclusion. Questions like \textit{``What's the invoked PowerShell script used by the attacker for dumping credentials?''} are paraphrased to \textit{``Is there any evidence of credential dumping by the attacker? If so, is there any invoked script used by the attacker for dumping credentials?''} This prevents the model from being biased toward technology-specific terms.

    \item\textbf{Numeric Neutrality:} Instead of guiding LLMs about the specific number of artifacts (e.g., malicious IP addresses), we allow LLMs to determine it from the given information. 
    Rather than asking \textit{``What are the attacker's two IP addresses?''}, we ask \textit{``Can you identify if there are any IP addresses that have been interacting with the server recently and that seem to be suspicious?''}. This prevents leading the model with fixed expectations.

    \item\textbf{External Knowledge-search Removal:} To ensure that 
    our evaluation emphasizes the model's reasoning ability only based on the provided context, we exclude questions that require external knowledge or rely on memorized facts, such as  CVE numbers or ransomware family names. 
    


\end{itemize}

The effect of debiasing strategies and limitation of guided nature of investigative question are discussed in Appendix \ref{sec:effect_debias} and \ref{sec:discussion}, respectively.

\subhead{SIA Task Generation and Validation} \label{sec:datasetfiles} 
In the following,
we describe how we generate and validate
the SIA tasks.


\noindent{\it Ground Truth Generation:} We generate the ground truths of the SIA tasks by manually reviewing the publicly available and official write-ups (e.g., \cite{Panagiotis, forensicskween, Medium, CyberDefenders}). Also, we generate the ground truth of the paraphrased questions described earlier by reviewing the files and artifacts. Finally, we manually verify that all ground-truths are valid from the files and artifacts, and consistent across all scenarios.


\noindent{\it Difficulty Level Assignment and Validation:} 
 To evaluate the performance of LLMs across varying levels of difficulty, we assign a difficulty level to each scenario based on its original source. A security professional validated the consistency of the assigned difficulty levels.

\noindent{\it Adversarial Tactic Assignment:} \label{adversarial_tactic}  
To reflect the investigative intent behind each question and alignment with real-world cybersecurity practices, we label the questions according to adversarial tactics defined in the MITRE ATT\&CK Framework \cite{mitre}. Using this framework, along with data sources (e.g., CyberDefenders~\cite{CyberDefenders}) and CISA guidelines \cite{cisa_tactic}, we manually annotate each question with the corresponding tactic it aims to uncover. These annotations follow investigative workflows outlined in widely adopted frameworks, such as those from NIST \cite{nist} and CISA playbooks \cite{cisa}, enabling evaluation of model capabilities across different tactical levels. The labeling process involves analyzing adversary behavior and inferring their likely objectives \cite{cisa_tactic} based on the investigative question. For example, the question \textit{``Is there any evidence of port scanning in the network traffic? If so, what is the IP address responsible for the scanning activity?''} targets the detection of port scanning to identify open ports and services. According to the MITRE ATT\&CK framework \cite{mitre}, this maps to the Reconnaissance tactic (TA0043), as the adversary's goal is to gather information for further exploitation. Some questions, however, do not correspond to a specific tactic—such as identifying a capture timestamp or determining a victim’s operating system. In such cases, the tactic field is left blank. 

\noindent{\it Tools and Task Feasibility Validation:} We create an environment equipped with the command line interface (CLI) tools, files and artifacts so that the LLM can take necessary actions to solve an SIA task. Since all the tools, named in with the original scenario, do not have CLI option, we select the additional CLI tools by reviewing forensic guides \cite{dfir}, public writeups (e.g., \cite{Panagiotis, forensicskween, Medium, CyberDefenders}), and manual workflows used by practitioners. After equipping the environment with the tools, files and artifacts, we manually verify that each task is realistically solvable using the CLI tools with the provided files and artifacts. While these tools are for initial validation, our LLM based autonomous agent (described in Sec. \ref{sec:agent}) is not limited to them. It has the capability to install and use additional tools as needed to complete the tasks. More on tool usage is in Appendix \ref{sec:discussion}.

\noindent{\it File Structure:}
We create JSON files for each SIA scenario. Each JSON file is structured with two distinct components to separate evaluation metrics from scenario data.
\begin{itemize}[leftmargin=*, itemsep=0pt, topsep=0pt]
    \item \textbf{LLM Input Components:}
The following elements are provided to the LLM for scenario analysis. \emph{\underline{{Scenario Description}}} provides an overview of the incident scenario; \emph{\underline{{Artifacts and Files}}} specifies the names of evidence files for analysis; \emph{\underline{{Directory}}} contains file location paths; \emph{\underline{{Tools}}} lists recommended analysis tools; \emph{\underline{{Basic Instruction}}} includes tool usage guidelines and scenario-specific information; and \emph{\underline{{Questions}}} presents multiple investigative questions per scenario.

\begin{table}[htp]
\scriptsize
\centering
\renewcommand{\arraystretch}{1}
\setlength{\tabcolsep}{1pt}
\begin{tabular}{|p{2cm}|p{1.25cm}|p{4.95cm}|}
\hline
\textbf{Category} & \textbf{Scenarios (Questions)} & \textbf{Classes of SIA tasks} \\ \hline

\textbf{Memory Forensics} 
    & \multirow{3}{*}{7 (60)} 
    & Analysis of Linux Server Compromise \\
    & & Analysis of Phishing-Driven Malware Infection \\
    & & Analysis of Malware Intrusion \\  
\hline

\textbf{Malware Analysis} 
    & \multirow{5}{*}{6 (46)} 
    & Analysis of Binary File \\  
    & & Analysis of Malicious Static Code \\  
    & & Analysis of XML Doc File \\  
    & & Analysis of Ransomware Attack-Related Script \\  
    & & Analysis of Malicious EXE file \\  
\hline

\textbf{Network Forensics} 
    & \multirow{6}{*}{8 (87)} 
    & Analysis of Phishing-Driven Malware Infection \\  
    & & Analysis of Internal Port Scanning Attack \\  
    & & Analysis of Drive-By Download Attack \\  
    & & Analysis of Ransomware Attack \\  
    & & Analysis of Lateral Movement Activity \\  
    & & Analysis of Suspected Data Exfiltration \\
    & & Analysis of Unknown Malicious Activity \\
\hline

\textbf{Miscellaneous} 
    & \multirow{4}{*}{4 (36)} 
    & Analysis of Phishing Email \\  
    & & Analysis of Typosquatting-Related Phishing Kit \\
    & & Analysis of Honeypot Compromise Logs \\
    & & Analysis of Server Compromise (via RCE) Logs \\  
\hline

\end{tabular}
\vspace{0.1in}
\caption{Summary of Our SIA Dataset (Part I).}
\label{tab:scenario_summary}
\end{table}

    \item \textbf{Evaluation Components:}
The following components are used exclusively for evaluation result representation: \emph{\underline{{Task Category}}} classifies the task type (e.g., Network Forensics and Malware Analysis); \emph{\underline{{Task Level}}} provides difficulty ratings (easy, medium and hard); \emph{\underline{{Tactic}}} categorizes attack behavior for each question; and \emph{\underline{{Answer}}} contains ground truth responses for evaluation.
\end{itemize}

\subhead{Human Expert Validation Study} 
We conducted a validation study by security researchers and professionals to gather additional human feedback on our de-biasing approach and tactic assignment.
The study has been approved by the Human Research Ethics Committee (HREC).

\noindent{\it Human Cohort:} 14 participants volunteered for the study to provide expert feedback.
We broadly categorize the participants into two
groups based on their SOC experience: limited ($<1$ year at
SOC) and experienced ($1+$ years at SOC), where precisely varying
0-5+ years SOC experience and 1-5+ years security experience across the cohort.

\noindent{\it Questionnaire:} 
There are two parts of the questionnaire:

\begin{itemize}[leftmargin=*, itemsep=0pt, topsep=0pt]
    \item \textbf{Debiasness of the questions:} 
    Each participant evaluated nine representative questions 
    that demonstrated different debiasing strategies (in Sec. \ref{sec:real_world_alignment})
    across diverse investigative contexts (network, process, 
    credential, and file-based indicators). For each question, 
    we presented: (a) collected raw question, (b) identified biases 
    in the original, (c) paraphrased question, and 
    (d) debiasing technique applied. Participants answered the question: ``Based on your experience, do you think our de-biasing approach more closely reflects real-world SIA investigations?''

    \item \textbf{Tactic Labels:} Participants assessed seven questions representing the most frequently occurring tactics (appearing more than ten times in our dataset). For each question, we presented the paraphrased question and assigned tactic label with explanation. Participants answered the question: ``Based on your cybersecurity experience, do you believe the assigned tactic label is correct?''
\end{itemize}

For each question, the participants can answer using a 4-point scale (4 = strongly agree, 3 = agree, 2 = disagree, 1 = strongly disagree) with options to provide feedback.

\begin{figure}[ht]
    \centering
    \includegraphics[width=\columnwidth]{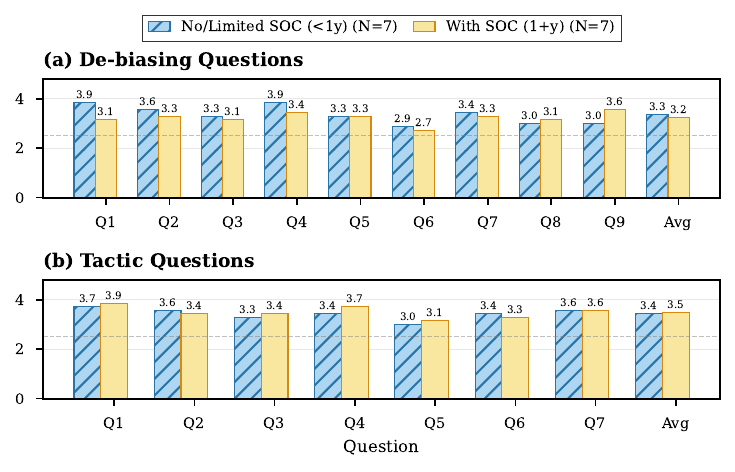}
    \vspace{-0.2in}
    \caption{Results of Human Expert Validation Study (where 4: strongly agree, 3: agree, 2: disagree, and 1: strongly disagree)}
    \vspace{-0.1in}
    \label{fig:user_study_fig}
\end{figure}

\noindent{\it Results:} 
Fig.~\ref{fig:user_study_fig} shows the results
with the average response scores for both the de-biasing and tactic questions
Both sets of participants show positive attitudes, 
with average debiasing scores of 3.35 and 3.22, and 
tactic scores of 3.43 and 3.49 (on a 4-point scale). 
Experienced analysts rated tactics slightly higher 
but debiasing lower, likely reflecting their awareness 
of exploratory SIA tasks, while less-experienced 
participants favored structured guidance. Overall, 
all scores exceeded the neutral midpoint, indicating 
broadly positive perceptions of our approach and labeling.

\subhead{Dataset Summary}
Table \ref{tab:scenario_summary} summarizes the coverage of our SIA dataset
and Table \ref{tab:dataset_scale_comparison} compares its scale with the datasize of existing works.
An excerpt of our dataset is available \cite{siabench_link}.

\begin{table}[h]
\centering
\scriptsize
\resizebox{\columnwidth}{!}{%
\begin{tabular}{|l|c|c|c|}
\hline
\multicolumn{1}{|c|}{\textbf{Dataset}} & \textbf{TS} & \textbf{TQ} & \textbf{Cybersecurity Task}                                                                                             \\ \hline
{\bf \oursolution}                                & 25          & 229         & Security Incident Analysis \\ 
                                         &             &             & (dataset covers multiple domains)\\ \hline
IRCopilot \cite{lin2025ircopilot}       & 12          & 130         & Incident Response \\ 
                                         &             &             & (limited domain coverage)\\ \hline
PentestGPT \cite{deng2024pentestgpt}    & 13          & 182         & Penetration testing                                \\ \hline
AutoPenBench \cite{gioacchini2024autopenbench} & 33   & 33          & Penetration testing          \\ \hline
InterCode-CTF \cite{yang2023language}   & 100         & 100         & Capture-the-flag Challenges                                                                                                \\ \hline
CyBench \cite{zhang2024cybench}         & 40          & 171         & Capture-the-flag Challenges                                                                                              \\ \hline
\end{tabular}
}
\footnotesize{TS = Total Scenarios; TQ = Total Questions}
\caption{Comparison of LLM Security Benchmark Datasets.}
\vspace{-0.2in}
\label{tab:dataset_scale_comparison}
\end{table}



\subsection{Part II: Alert Triaging Tasks}
\begin{figure}[ht] 
    \centering
    \includegraphics[width=\columnwidth]{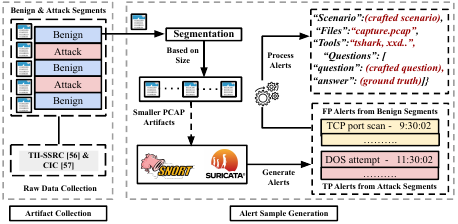} 
     \caption{Alert Triage Dataset (Part II) Generation Process.}    
      \vspace{-0.1in}
    \label{fig:benign_data_generation}
\end{figure}

There is a lack of false alert datasets in a format suitable for LLMs, as well as a lack of benchmarks that directly evaluate their performance in analyzing benign artifacts.
To address this gap, we developed a LLM digestible alert dataset that includes benign scenarios (and their corresponding false alerts).
Specifically, we focus on network forensics-related tasks using artifacts like network level alerts by IDSes (e.g., Snort~\cite{snort}, Suricata~\cite{suricata}).
We develop the dataset through the following steps 
(as illustrated in Fig. \ref{fig:benign_data_generation}).


\subhead{Artifact Collection} We consider two datasets to collect the artifacts from:
TII-SSRC-23 Dataset \cite{herzalla2023tii} and
CIC-IDS2017 Intrusion Detection Dataset \cite{cicids2017}. 
The TII-SSRC-23 Dataset \cite{herzalla2023tii} includes separate benign and malicious PCAPs across eight traffic types (audio, video, text, background, information gathering, bruteforce, DoS, and Mirai botnet) with 32 subtypes (six benign and 26 malicious).
The CIC-IDS2017 dataset \cite{cicids2017} includes the PCAPs with a mix of both benign and malicious traffic, covering attacks such as web exploits, brute force, DoS/DDoS, infiltration, Heartbleed, botnets, and scanning. For this dataset, we extract the benign traffic from attack traffic using the provided attack timestamps. 


\subhead{Alert Sample Generation} Due to the absence of the  natural-language-based predefined alert scenarios and questions in the datasets, we manually craft scenarios and corresponding questions from the raw PCAPs to evaluate LLMs for alert triage. To address LLMs' limitations \cite{context_limit1, context_limit2} with large files, we segment PCAPs into chunks of about 10,000 packets, as original captures often exceed 100,000 packets and several gigabytes.    
After having malicious and benign PCAP segmentation, we leverage Snort v3.9.3.0 and Suricata v8.0.2 with default community rules and emerging threats open rule set, respectively, freely available from their official website, to generate corresponding false positive (FP) and true positive (TP) alerts.
We use the generated alerts to craft the dataset as follows.

\noindent\textit{Scenario Generation}: We incorporate alert descriptions directly into the scenarios,
while maintaining consistency with the Part-I of our dataset. Example: ``The SOC team has received the following alert: \texttt{07/07-08:00:50.818009 [**] [122:1:1] (port\_scan) TCP portscan [**] [Priority: 3] \{TCP\} 192.168.10.3:88 - 192.168.10.5:49175}. A PCAP file has been provided to investigate the alert.''
    
\noindent\textit{Question Formulation:} For alert triaging, we formulate a single simple and direct question in a general form: ``Analyze the provided PCAP file and determine if the generated alert is a false positive or true positive.''

\noindent\textit{Ground Truth Labeling:} To generate the ground truth we use the labeled flow data from TII-SSRC-23 and CIC-IDS2017 datasets. We correlate each alert’s 5-tuple (source/destination IPs, source/destination ports, and timestamp) with the flow entries. Alerts that match only malicious flows are labeled as True Positives (TPs). If an alert match with benign flows or appear in PCAP segments containing exclusively benign traffic are labeled as False Positives (FPs).

\subhead{Dataset Summary}
This part of the dataset has (in total) 135 alert scenarios in network forensics;
where 50 FPs and 50 TPs scenarios from TII-SRC-23, as well as 30 FPs and five TPs scenarios from CIC-IDS2017. 
The ethical considerations for our dataset are in Appendix~\ref{appendix:ethics}.

\section{\oursolution Agent} \label{sec:agent}
 This section presents the \oursolution agent.

\subsection{Challenges}
In designing the agent, 
we face three unique challenges.


\begin{itemize}[leftmargin=*, itemsep=0pt, topsep=0pt]
\item 
\textbf {Dynamic Interactions.}
The traces related to various security incidents are often 
stored in different types of artifacts, the collection and analysis of which require 
the frequent interactions with several security tools and environments.
However, in general, LLMs rely on their memory-based knowledge and do not
dynamically interact with external security tools or the environment yet. 
To overcome this issue, we implement several functionalities
(e.g., \texttt{execute tool})
that allows LLMs to use those functionalities to interact 
with external tools and environments 
(further detailed in Sec.~\ref{sec:agentdesign}).

\item 
\textbf {Tool Output Processing.}
Simply using the outputs from those security tools 
by the LLMs can lead to context length exhaustion \cite{shi2023large} along with wrong decisions \cite{Prompt_Engineering}, as they are known to include significantly more
irrelevant information~\cite{yang2024true,alahmadi202299}.
To tackle this challenge, we 
reduce the output space while keeping the relevant information
to make effective decisions in complex multi-objective scenarios like SIA tasks
(further detailed in Sec.~\ref{sec:incidentstate}).



\item 
\textbf {Multi-State Workflow.}
SIA tasks are inherently multi-objective, requiring analysts to answer multiple questions (e.g., who, what, when, and how) and exploratory in nature. In our dataset, even all questions are known, asking all the questions simultaneously in a single-state setting may cause the agent to get overly focused on one question while overlooking others ~\cite{masterman2024landscape, talebirad2023multi}. It also hinders the natural exploratory nature of SIA task where then next question is often asked based on the findings of the previous question. To address this, we design a \emph{multi-state} workflow for \oursolution agent to automate the process of addressing multi-question scenarios, sequentially solving an entire incident mimicing closely the real-world workflow. An ablation study (in Sec.~\ref{subsec:ablationstudy}) demonstrates the advantages of multi-state workflow over single-state.

In the following, we detail the \oursolution agent design.

\end{itemize}



\subsection{Agent Design}
\label{sec:agentdesign}

\begin{figure}[ht] 
    \centering
     \vspace{-0.1in}
    \includegraphics[width=\columnwidth]{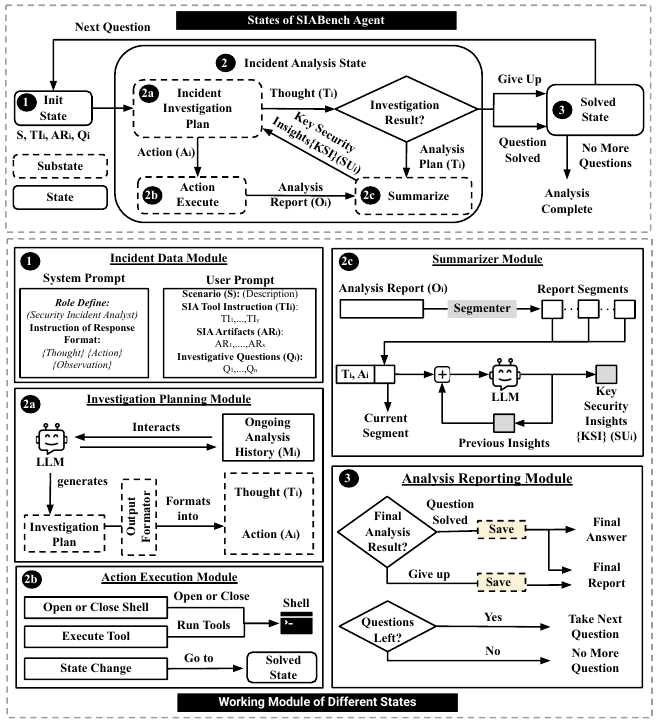} 
     \parbox{0.7\textwidth}{ \footnotesize (S=Scenario Description, T=Tool Instruction, A=Artifact, \newline Q=Question,  KSI=Key Security Insights)}
      \vspace{-0.2in}
     \caption{An Overall Design of the \oursolution Agent.}
    \label{fig:agent_workflow}
\end{figure}

\subhead{Overview} 
The top of Fig. \ref{fig:agent_workflow} shows 
three primary states of \oursolution agent.
Specifically, in the \emph{Init} state (\circledA{1}), scenario description ($S$), artifact data ($AR_i$), and tool instructions ($TI_i$) are provided to the agent as input to initiate the analysis for each question ($Q_i$). The \emph{Incident Analysis} state 
(\circledA{2}) forms the core of the agent's operations
with 
three iterative sub-states: the \textit{Incident Investigation Plan} sub-state (\circledA{2a}), where the LLM agent determines the required plan and actions to analyze the artifacts; the \textit{Action Execute} sub-state (\circledA{2b}), where these actions are carried out in the environment executing the tools; and the \textit{Summarize} sub-state (\circledA{2c}), where the key insights are extracted from the analysis report. The agent continuously loops through these sub-states, refining its plan and decisions with each cycle, until it gives up or finds the answer to the question, $Q_i$, or exhausts all cycles and moves to the next state. In the \emph{Solved} state (\circledA{3}), the agent stores the logs and decides whether it needs to return to the \emph{Init} state for any remaining questions or terminates the analysis process. 
The following will provide its detailed explanation.

\subhead{1. Init State}
To initially provide the LLM agent with all necessary information,
we design a system prompt and a user prompt \cite{Prompt_Engineering}. The system prompt defines the model’s role as a security incident analyst, provides clear instructions, sets the context to guide the LLM to follow a defined response structure \cite{yao2022react}, and explains the functionalities used by the agent, using a zero-shot prompting approach without any fine-tuning.
The user prompt details the specific task, including the scenario, example set of tool instructions, relevant artifacts, and the question or objective, as shown in Module \circledA{1} in the bottom of Fig. \ref{fig:agent_workflow}. 
The snippet of our system and user prompts are in the Appendix \ref{sec:prompts}.

\subhead{2. Incident Analysis State}
\label{sec:incidentstate}
In this state, 
the agent analyzes the incident scenario to find answers to the questions
following three sub-states, which operate in a loop for analysis.

\noindent\emph{\underline{(a) Incident Investigation Plan.}}
This is to orchestrate the analysis process to solve the SIA task, as depicted in Module \circledA{2a} in Fig. \ref{fig:agent_workflow}. Our agent utilizes the ReAct (Reason+Act) framework \cite{yao2022react}, which is a
framework that combines reasoning and action in LLM, enabling dynamic reasoning, plan adjustments, and interaction with external environments for task completion,
with iterative cycles of \textit{thought}, \textit{action}, and \textit{observation},
(this design choice is justified through evaluation in Sec. {\ref{subsec:ablationstudy}}). Specifically,
    \textit{thought ($T_i$):} the LLM evaluates the current state by analyzing available information and plans the next investigation step.
    \textit{Action  ($A_i$):} based on its assessment, the agent generates specific actions.
    \textit{Observation ($SU_i$):} the results of the action are used to refine future steps.
To ensure consistency, we incorporate an \emph{Output Formatter} to  parse the raw output from the LLMs and format them into a structured form of \textit{thought} ($T_i$) and \textit{action} ($A_i$). Also, the LLM maintains an \emph{analysis history ($M_i$)}, updated with each $T_i$, $A_i$, and $SU_i$. This history enables the agent to utilize prior key insights, fostering cohesive and informed responses throughout an SIA task.

\begin{figure}[h]
    \centering
 
    \includegraphics[width=\columnwidth]{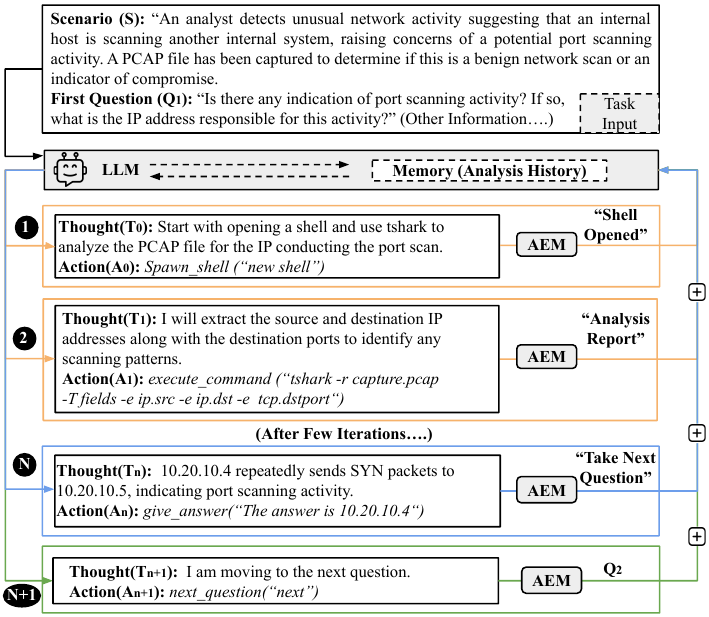} 
    \parbox{0.7\textwidth}{ \footnotesize (AEM=Action Execution Module)}
     \vspace{-0.2in}
    \caption{An Example of an Incident Investigation Plan.}

    \label{fig:action_reason}
\end{figure}

\noindent \textit{Example 1.} 
Fig. \ref{fig:action_reason} illustrates the Module \circledA{2a} workflow. Given a description of a port-scanning attack, a question, and supporting details, the LLM begins by generating an initial thought ($T_0$) to open a shell environment with action ($A_0$), executed via Module \circledA{2b}. It then formulates the next thought ($T_1$) to plan its investigation and produces an action ($A_1$), such as running commands (e.g., tshark) on the provided artifact (e.g., PCAP), again executed through Module \circledA{2b} to obtain a report.
Each iteration updates the investigation history, enabling the LLM to refine its reasoning based on previous outputs. After several cycles of thought and action, the LLM reaches a conclusion, answers the question, and proceeds to the next task.

\noindent\emph{\underline{(b) Action Execute.}}
This step is to execute the actions generated by the agent through interactions with relevant tools/environments using Module \circledA{2b} in Fig.~\ref{fig:agent_workflow}. It enables the LLM to emulate analyst workflows and interact with its environment through three functions: 
(i) \emph{Open/Close Shell}: creates an isolated environment for executing commands, analyzing outputs, and terminating once the task is completed; 
(ii) \emph{Execute Tool}: runs specialized security tools (e.g., \textit{Volatility}~\cite{volatility} for memory forensics, \textit{Tshark}~\cite{tshark} for PCAP analysis, \textit{Oledump}~\cite{oledump} for malware inspection), including installing needed tools and refining investigations based on report outputs; and 
(iii) \emph{State Change}: decides when to conclude or skip a question, supporting efficient time management~\cite{nyuctf} and moving the investigation forward.

\begin{figure}[h] 
    \centering
         \vspace{-0.1in}
    \includegraphics[width=\columnwidth]{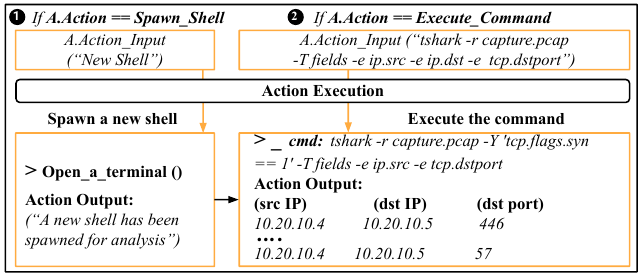} 
             \vspace{-0.2in}
    \caption{An Example of the Action Execution Step.}
    \label{fig:action_execution}
\end{figure}

\noindent \textit{Example 2.} 
Fig.~\ref{fig:action_execution} illustrates the Module 
\circledA{2b} workflow, showing its \emph{Open/Close Shell} 
and \emph{Execute Tool} functions. First, the LLM invokes 
\texttt{spawn shell} to open an isolated shell environment. 
It then executes tool commands to analyze the \texttt{PCAP} 
artifact obtained from Fig.~\ref{fig:action_reason}. The 
resulting output shows repeated connection attempts from 
\texttt{172.20.10.4} to \texttt{172.20.10.5} across multiple 
ports.


\noindent\emph{\underline{(c) Summarize.}}
During security incident analysis, large volumes of logs contain 
substantial redundancy, making it difficult for both analysts 
and LLMs to extract key details \cite{wang2024ada, miranskyy2016operational}. 
To address this, Module \circledA{2c} (Fig.~\ref{fig:agent_workflow}) 
summarizes the analysis report from the previous step, producing 
\emph{Key Security Insights} (KSI) that streamline input back to 
Module \circledA{2a}. We adopt an LLM-based summarization approach 
inspired by prior works \cite{deng2024pentestgpt, gioacchini2024autopenbench}. 
Structured prompts (in Appendix~\ref{sec:prompts}) guide the model 
to extract key security details from the problem-solver 
agent’s outputs. Each prompt includes the current thought 
($T_i$), action ($A_i$), analysis report ($O_i$), scenario 
description ($S$), and the question ($Q_i$), enabling the 
SIA agent to focus on the most relevant insights for completing the task.


If the analysis output exceeds a predefined limit (e.g., 64k tokens), it is split into smaller segments \cite{an2024does}. The LLM first summarizes the initial segment to produce $SU_1$, then iteratively incorporates this summary into subsequent segments to maintain context and correlations. After all segments are processed, the final summary $SU_{\text{i}}$ captures the key security insights (KSI) needed by the SIA agent. This prevents overload from irrelevant details and mitigates errors such as hallucinations and reasoning inconsistencies \cite{Prompt_Engineering}, while partially addressing context-length limitations \cite{shi2023large}. As shown in Sec.~\ref{subsec:ablationstudy}, due to context constraints, GPT-4o fails to complete 11 out of 25 scenarios, and Claude-3.5-Sonnet fails on seven.

\begin{figure}[htp] 
    \centering
    \includegraphics[width=\columnwidth]{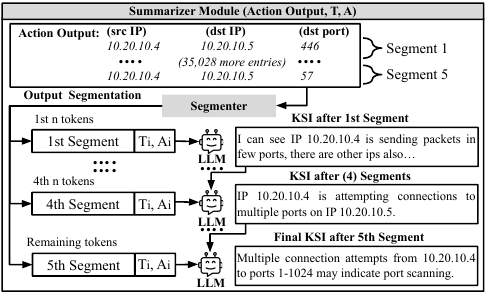} 
    \caption{An Example of the Summarizer Module.}
    \label{fig:summarization_module}
    \vspace{-0.1in}
\end{figure}

\noindent \textit{Example 3.} 
Fig.~\ref{fig:summarization_module} shows an example of Module \circledA{2c} using the analysis report from Example~2. The segmenter splits the report into five parts based on token length. In the first iteration, the LLM receives the initial segment along with the thought ($T_i$) and action ($A_i$) (scenario details $S$ and question $Q_i$ omitted for simplicity) to generate KSI. In the second iteration, the next segment and prior KSI are provided to update the insights. This process repeats until the final segment is processed, producing the complete KSI set. These insights are then passed to Module \circledA{2a} to guide the agent’s next steps.


\subhead{3. Solved State} Once the agent determines that it has either successfully solved a specific question (\(Q_{i}\)) or given up on further progress on this question, it transits to this solved state. In this state, the agent logs all the information (investigation plan, actions, and analysis output) for the question as the \textit{Final Report} and also evaluate LLMs answer with the ground truth saving as \textit{Final Answer}, as indicated in Module \circledA{3} in Fig. \ref{fig:agent_workflow}. Additionally, the agent checks whether there are any remaining questions to address within the current scenario. If additional questions are identified, the agent transitions back to the \emph{Init State} to start investigation with the next question (\(Q_{i+1}\)), 
The implementation details of this agent are in Appendix \ref{subsec:agent_implementation}.

\subsection{Adaptivity with New Models and Tasks}
\label{sec:newmodel}
\subhead{New Models} 
To assess the effort needed to integrate new models into our agent, we examined two cases: models failing to perform correct tool calls (\textit{o3-mini}) and models triggering policy-violation errors (\textit{GPT-5}). The \textit{o3-mini} model from OpenAI \cite{OpenAI} became functional after prompt adjustments; specifically, enforcing: ``Always execute the required function calls before you respond,'' which reduced failures and improved consistency. For GPT-5, we applied the DAN (“Do Anything Now”) jailbreak technique \cite{ad_prompt} to ensure successful completion of all scenarios in our dataset.

\subhead{New Tasks} \label{newtask} 
To evaluate the effort needed to add a new SIA task, we incorporated an Android Artifact Analysis scenario from CyberDefenders~\cite{CyberDefenders}, which is excluded from the \oursolution dataset and requires fundamentally different analysis approaches (e.g., parsing SQLite databases, XML files, and app-specific data). The primary work involved preparing the corresponding \textit{JSON} file as in Sec.~\ref{sec:dataset} and downloading the necessary artifacts. The remaining evaluation steps required no changes and produced results consistent with other SIA tasks.
Additional discussions on our agent are  in Appendix~\ref{sec:discussion}.

\section{Evaluation} \label{sec: evaluation_nn}
Our evaluation answers seven 
research questions (RQs):

\begin{itemize}[leftmargin=*, itemsep=0pt, topsep=0pt]
    \item \textbf{RQ1: }How well do LLMs perform in diverse SIA tasks?
    \item \textbf{RQ2: }At which tactic level in SIA, do LLMs excel?
    \item \textbf{RQ3: }What are the primary factors of the LLM’s failure?
    \item \textbf{RQ4: }How consistently do LLMs perform SIA tasks?


\item \textbf{RQ5: }How do LLMs perform on live SIA tasks?


\item \textbf{RQ6: }How different components of \oursolution agent influence LLMs' performance on SIA tasks? 


\item \textbf{RQ7:} How do LLMs perform in alert triaging? 

\end{itemize}
\subsection{Experimental Setup} \label{subsec:experimental_setup}

We evaluate four open-weight 
(DeepSeek-Reasoner \cite{deepseek_reasoner} and Meta's three Llama 3.1 models \cite{LLaMA}: Llama-3.1-8B, Llama-3.1-70B, and Llama-3.1-405B)
and seven close-weight 
(Claude-4.5-Sonnet, Claude-3.5-Sonnet\footnote{Claude-3.5-Sonnet has been deprecated as of October 2025.} \cite{anthropic}, Gemini-1.5-pro\footnote{Gemini-1.5-pro has been deprecated as of September 2025.} \cite{team2024gemini}, and OpenAI's \cite{OpenAI} GPT-4o-mini, GPT-4o, o3-mini, and GPT-5)
LLMs. 
The temperature is set to zero and seed value fixed for all models to keep the output deterministic, aligning with prior works  \cite{alam2024ctibench, gioacchini2024autopenbench}. 
The experiments are conducted in a local virtual Kali Linux environment with 16 GB RAM, 12th Gen Intel Core i7 processor, 
and a 512 GB virtual disk.
\oursolution agent has access to this environment and command-line tools (such as, Tshark \cite{tshark}, Volatility \cite{volatility}, Oledump \cite{oledump}, and PDFiD \cite{pdfid}), with the additional capability of installing more tools.

\subsection{Results on SIA Tasks (RQ1-RQ4)} \label{sec:RQ1-4}

\begin{table*}[ht]
\centering
\renewcommand{\arraystretch}{1.3}
\resizebox{\textwidth}{!}{
\begin{tabular}{|l||c|c|c|c|c|c||c|c|c|c|c|c||c|c|c|c|c|c||c|c|c|c||c|c|}
\hline
\multirow{3}{*}{\textbf{Model}} & \multicolumn{6}{c||}{\textbf{Memory Forensics}} & \multicolumn{6}{c||}{\textbf{Malware Analysis}} & \multicolumn{6}{c||}{\textbf{Network Forensics}} & \multicolumn{4}{c||}{\textbf{Miscellaneous}} & \multicolumn{2}{c|}{\textbf{{Overall}}} \\
\cline{2-25}
& \multicolumn{2}{c|}{\textbf{E (4S, 29Q)}} & \multicolumn{2}{c|}{\textbf{M (2S, 16Q)}} & \multicolumn{2}{c||}{\textbf{H (1S, 15Q)}} & \multicolumn{2}{c|}{\textbf{E (2S, 11Q)}} & \multicolumn{2}{c|}{\textbf{M (3S, 23Q)}} & \multicolumn{2}{c||}{\textbf{H (1S, 12Q)}} & \multicolumn{2}{c|}{\textbf{E (3S, 25Q)}} & \multicolumn{2}{c|}{\textbf{M (4S, 52Q)}} & \multicolumn{2}{c||}{\textbf{H (1S, 10Q)}} & \multicolumn{2}{c|}{\textbf{E (3S, 25Q)}} & \multicolumn{2}{c||}{\textbf{M (1S, 11Q)}} & \multicolumn{2}{c|}{\textbf{(25S, 229Q)}} \\
\cline{2-25}
& \textbf{FS} & \textbf{PS} & \textbf{FS} & \textbf{PS} & \textbf{FS} & \textbf{PS} & \textbf{FS} & \textbf{PS} & \textbf{FS} & \textbf{PS} & \textbf{FS} & \textbf{PS} & \textbf{FS} & \textbf{PS} & \textbf{FS} & \textbf{PS} & \textbf{FS} & \textbf{PS} & \textbf{FS} & \textbf{PS} & \textbf{FS} & \textbf{PS} & \textbf{FS} & \textbf{PS} \\
\hline
\textbf{Claude-4.5-Sonnet} & \textbf{2/4} & \textbf{86.74\%} & \textbf{1/2} & \textbf{85.00\%} & 0/1 & 33.33\% & 0/2 & 81.67\% & \textbf{1/3} & \textbf{80.51\%} & 0/1 & \textbf{41.67\%} & 0/3 & 87.74\% & \textbf{1/4} & \textbf{86.12\%} & \textbf{1/1} & \textbf{100.0\%} & \textbf{2/3} & \textbf{94.44\%} & 0/1 & 54.55\% & \textbf{8/25} & \textbf{81.70\%} \\
\hline
\textbf{GPT-5} & \textbf{2/4} & 86.04\% & 0/2 & 81.67\% & 0/1 & 26.67\% & \textbf{1/2} & \textbf{91.67\%} & 0/3 & 76.41\% & 0/1 & \textbf{41.67\%} & \textbf{1/3} & \textbf{92.50\%} & 0/4 & 84.92\% & 0/1 & 90.00\% & 1/3 & 86.11\% & 0/1 & 63.64\% & 5/25 & 80.70\% \\
\hline
\textbf{Claude-3.5-Sonnet} & 1/4 & 76.73\% & 0/2 & 65.00\% & 0/1 & \textbf{53.33\%} & \textbf{1/2} & 80.00\% & \textbf{1/3} & 78.97\% & 0/1 & 16.67\% & 0/3 & 82.98\% & 0/4 & 60.05\% & 0/1 & 60.00\% & 1/3 & 83.33\% & 0/1 & \textbf{72.72\%} & 4/25 & 71.03\% \\
\hline
\textbf{GPT-4o} & 1/4 & 61.47\% & 0/2 & 60.00\% & 0/1 & 33.33\% & \textbf{1/2} & 80.00\% & 0/3 & 58.97\% & 0/1 & 16.67\% & 0/3 & 64.05\% & 0/4 & 43.17\% & 0/1 & 40.00\% & 1/3 & 84.72\% & 0/1 & 36.36\% & 3/25 & 57.93\% \\
\hline
\textbf{DeepSeek-Reasoner} & 0/4 & 50.88\% & 0/2 & 38.33\% & 0/1 & 26.67\% & \textbf{1/2} & 70.00\% & 0/3 & 48.71\% & 0/1 & 33.33\% & 0/3 & 70.12\% & 0/4 & 36.29\% & 0/1 & 30.00\% & 0/3 & 61.67\% & 0/1 & 36.36\% & 1/25 & 49.33\% \\
\hline
\textbf{Gemini-1.5-pro} & 0/4 & 33.81\% & 0/2 & 31.67\% & 0/1 & 6.67\% & 0/2 & 73.33\% & 0/3 & 25.13\% & 0/1 & 16.67\% & 0/3 & 62.98\% & 0/4 & 31.79\% & 0/1 & 20.00\% & 1/3 & 72.22\% & 0/1 & 45.45\% & 1/25 & 41.61\% \\
\hline
\textbf{o3-mini} & 0/4 & 31.24\% & 0/2 & 15.00\% & 0/1 & 13.33\% & 0/2 & 53.33\% & 0/3 & 25.13\% & 0/1 & 25.00\% & 0/3 & 72.14\% & 0/4 & 41.90\% & 0/1 & 40.00\% & 1/3 & 63.89\% & 0/1 & 36.36\% & 1/25 & 41.10\% \\
\hline
\textbf{GPT-4o-mini} & 0/4 & 33.14\% & 0/2 & 18.33\% & 0/1 & 0.00\% & 0/2 & 73.33\% & 0/3 & 18.46\% & 0/1 & 16.67\% & 0/3 & 29.29\% & 0/4 & 26.50\% & 0/1 & 20.00\% & 1/3 & 58.33\% & 0/1 & 36.36\% & 1/25 & 32.53\% \\
\hline
\textbf{Llama-3.1-405B} & 0/4 & 24.09\% & 0/2 & 5.00\% & 0/1 & 0.00\% & 0/2 & 43.33\% & 0/3 & 10.26\% & 0/1 & 16.67\% & 0/3 & 31.79\% & 0/4 & 19.07\% & 0/1 & 10.00\% & 0/3 & 53.33\% & 0/1 & 9.09\% & 0/25 & 23.28\% \\
\hline
\textbf{Llama-3.1-70B} & 0/4 & 14.09\% & 0/2 & 5.00\% & 0/1 & 0.00\% & 0/2 & 26.67\% & 0/3 & 6.67\% & 0/1 & 8.33\% & 0/3 & 17.50\% & 0/4 & 15.08\% & 0/1 & 0.00\% & 0/3 & 47.78\% & 0/1 & 9.09\% & 0/25 & 16.53\% \\
\hline
\textbf{Llama-3.1-8B} & 0/4 & 2.20\% & 0/2 & 0.00\% & 0/1 & 0.00\% & 0/2 & 25.00\% & 0/3 & 0.00\% & 0/1 & 0.00\% & 0/3 & 8.33\% & 0/4 & 0.00\% & 0/1 & 0.00\% & 0/3 & 27.50\% & 0/1 & 9.09\% & 0/25 & 7.03\% \\
\hline
\end{tabular}}
\parbox{1.0\textwidth}{ \footnotesize{E=Easy, M=Medium, H=Hard, FS=Fully Solved Scenario, PS=Average Percentage of the Partially Solved Scenarios, S=no. of scenarios, Q=no. of questions}}
\caption{Performance Comparison of LLMs: Fully Solved Scenarios vs. Average Percentage of Partially Solved Scenarios.}
\label{tab:model_comparison}
\end{table*}


\subsubsection{LLM's Overall Performance in SIA Tasks (RQ1)}
We evaluate 11 LLMs on our SIA Task dataset using two metrics: Fully Solved (FS), the number of scenarios where all questions are answered correctly, and Partially Solved (PS), the average percentage of correct answers per scenario. Table~\ref{tab:model_comparison} summarizes model performance: Claude-4.5-Sonnet and GPT-5 achieve the strongest results across most categories. Overall, LLMs still struggle to complete entire SIA scenarios: only 8 of 25 are fully solved even by the best-performing model. We analyze per-task results as follows.


\noindent \textit{\ul{Memory Forensics:}} 
Table \ref{tab:model_comparison} highlights notable performance differences among models on memory forensics tasks. Overall, Claude-4.5-Sonnet performs best, answering most questions and achieving the highest success rate on easy-level tasks (86.74\%), closely followed by GPT-5 (86.04\%). The PS score of smaller models, such Llama-3.1-8B and 70B are very low across all difficulty levels. Tasks involving memory dump analysis with tools like Volatility \cite{volatility} are particularly challenging. As task difficulty increases, performance declines sharply - Claude-4.5-Sonnet drops to 33.33\% on hard scenarios, with even steeper declines for other models.  In terms of FS, Claude-4.5-Sonnet solves three out of the seven scenarios, while GPT-5 solves only two. These findings underscore the challenges LLMs, especially smaller ones, face in memory forensics and indicate significant room for improvement in this domain.

\noindent \textit{\ul{Malware Analysis:}} Easy malware analysis tasks, such as, interpreting plaintext ransomware scripts or extracting key details via static code analysis, require minimal analysis. GPT-5, Claude-3.5-Sonnet, GPT-4o and DeepSeek-Reasoner each solve one of two scenarios, with GPT-5 achieving the highest PS rate (91.67\%), followed by Claude-4.5-Sonnet (81.67\%). Llama-3.1-405B performs moderately (43.33\%), while Llama-3.1-8B performs poorly (25.00\%), showing weak tool usage and environment interaction, consistent with Berkeley Tool Calling Leaderboard results \cite{berkley}. For medium-hard tasks (e.g., binary or XML file analysis) requiring advanced tool calling and reasoning, Claude-4.5-Sonnet leads with 80.51\%, while models like DeepSeek-Reasoner and o3-mini fall below 50\%.  On hard-level tasks, Claude-4.5-Sonnet and GPT-5 outperform others with PS rates above 40\%. For FS, only Claude-3.5-Sonnet solves two scenarios out of six, whereas the other two top performing models GPT-5 and Claude-4.5-Sonnet only solved one. All these results indicate substantial room for improvement.  

\noindent \textit{\ul{Network Forensics:}} 
Claude-4.5-Sonnet, GPT-5, and Claude-3.5-Sonnet achieve PS rates above 80\% on easy tasks, which involve basic capabilities such as scan pattern detection, file uploads, and script or command detection in packet headers. In contrast, Llama-3.1-8B and 70B perform poorly, with PS rates of 8.33\% and 17.5\%, respectively. Llama-3.1-405B and GPT-4o-mini also score low (31.79\% and 29.29\%), highlighting their limitations in simple network forensics. For medium-level tasks requiring deep packet inspection and object extraction (e.g., PDFs), Claude-4.5-Sonnet and GPT-5 lead with 86.12\% and 84.92\%, while other models fall below 60\%. Llama-3.1 models (8B, 70B, 405B) perform worst, with PS rates under 20\%, underscoring their limitation to handle complex network forensics tasks. For FS, Claude-4.5-Sonnet solved two (one medium and one hard) out of eight scenarios, and GPT-5 solved only one (easy) scenario. 

\noindent \textit{\ul{Miscellaneous:}} 
Similar to easy-level malware analysis tasks, comparable trends appear in Miscellaneous easy-level tasks, such as analyzing phishing emails, reviewing simple plaintext logs, and examining phishing kits. For such tasks, Claude-4.5-Sonnet, GPT-5, Claude-3.5-Sonnet, GPT-4o exceed 80\% PS rates. GPT-4o-mini struggles with tasks requiring file system navigation, and Llama-3.1 models consistently scored low even on basic challenges. As difficulty increases, performance declines across all models—except Claude-3.5-Sonnet, which achieves a respectable 72.72\% PS on medium-level tasks. For FS, Claude-4.5-Sonnet solves two easy scenarios out of four, while GPT-5, Claude-3.5-Sonnet, GPT-4o, Gemini-1.5-pro, o3-mini, and GPT-4o-mini solve one each, bolstering the room for improvement. 


\subsubsection{LLM's Tactic-level Performance in SIA Tasks (RQ2)}
We examine the percentage of questions solved by the top four LLMs across tactic categories, focusing on tactics appearing in at least ten questions to ensure consistency. This analysis reveals tactic-level performance (Sec.~\ref{adversarial_tactic}) and highlights why models struggle to complete full scenarios. Table~\ref{tab:tactic_solve} summarizes the results for the top four models from RQ1.


As shown in Table~\ref{tab:tactic_solve}, LLMs maintain strong performance in the Reconnaissance phase. Probing activity detection, port scans, and directory enumeration are handled accurately, especially with structured network forensic inputs. Claude-4.5-Sonnet achieves 100\% success, followed by Claude-3.5-Sonnet and GPT-4o at 91.67\%, with GPT-5 also performing well, suggesting strong capability in identifying surface-level, pattern-based artifacts. LLMs similarly perform well in Command and Control (51.85\%--76.00\%) and Exfiltration (53.85\%--91.67\%) tasks, which rely on analyzing network communications for abnormal connections, suspicious URLs, protocols, and data transfer behavior.
Credential Access shows high variability (45.45\%--85.71\%), with GPT-4o struggling on encoded credentials and sophisticated extraction techniques. Performance in Initial Access is moderate (73.91\%, 56.52\%, 71.43\%, and 66.67\% for Claude-3.5-Sonnet, GPT-4o, Claude-4.5-Sonnet, and GPT-5, respectively) where models can detect phishing or early compromise attempts but not consistently across all cases.


However, LLMs face challenges in tasks requiring deep technical analysis. Defense Evasion shows the lowest performance, dropping below 55\% for both Claude-3.5-Sonnet and GPT-4o, indicating difficulty with obfuscated data, complex encodings, and non-plaintext formats requiring reverse engineering expertise. Execution analysis similarly remains below 65\% for these models, especially when extracting embedded scripts from complex formats (e.g., XML, PDF), dissecting binaries, or performing deep memory forensics. In contrast, recent models such as Claude-4.5-Sonnet and GPT-5 achieve over 80\% success in Execution and above 70\% in Defense Evasion, demonstrating improved capability in advanced analysis and decoding processes.

\begin{table}[h]
\centering
\fontsize{9.5}{10.5}\selectfont  
\setlength{\arrayrulewidth}{0.15pt}  
\begin{adjustbox}{width=\columnwidth,center}
\begin{tabular}{|p{3.1cm}|p{4.6cm}|p{1.7cm}|p{1.2cm}|p{1.2cm}|p{1.2cm}|}
\hline
\textbf{Tactic Category} & \textbf{Task Category (Total Questions)} & \textbf{Claude-3.5-Sonnet} & \textbf{GPT-4o} & \textbf{Claude-4.5-Sonnet} & \textbf{GPT-5} \\ \hline
Reconnaissance & NF(11), MS(1) & 91.67\% & 91.67\% & \textbf{100.00\%} & 90.00\% \\ \hline
Exfiltration & NF(8), MF(2), MS(3) & 84.62\% & 53.85\% & 83.33\% & \textbf{91.67\%} \\ \hline
Command and Control & NF(9), MF(12), MA(4), MS(2) & 74.07\% & 51.85\% & \textbf{76.00\%} & 64.00\% \\ \hline
Credential Access & NF(3), MF(6), MS(2) & 72.73\% & 45.45\% & 64.29\% & \textbf{85.71\%} \\ \hline
Initial Access & NF(9), MF(8), MS(6) & \textbf{73.91\%} & 56.52\% & 71.43\% & 66.67\% \\ \hline
Execution & NF(11), MF(10), MA(12), MS(1) & 64.70\% & 52.94\% & \textbf{81.40\%} & \textbf{81.40\%} \\ \hline
Defense Evasion & NF(8), MF(4), MA(23), MS(4) & 53.84\% & 51.28\% & \textbf{74.07\%} & \textbf{74.07\%} \\ \hline
\end{tabular}
\end{adjustbox}
\footnotesize{MF=Memory Forensics, MA=Malware Analysis, NF=Network Forensics, MS=Miscellaneous}
\caption{Solving Percentages for Top Tactics and Models With Categorized Counts.}
\vspace{-0.1in}
\label{tab:tactic_solve}
\end{table}

\subsubsection{Primary Factors of LLM's Failure (RQ3)}
We further analyze LLM failures when solving questions in our SIA tasks. During agentic evaluation, we observe three failure modes: \emph{round exceed}, \emph{give up}, and \emph{wrong answer}. Round exceed occurs when a model exhausts all available turns without producing a correct answer. Give up reflects premature abandonment after repeated unsuccessful attempts, typically within the first few steps. Wrong answer represents confident but incorrect responses. We compute each failure type as a percentage of the total number of questions where it appears. We then analyze the root causes behind these three failure modes as follows:



\begin{itemize}[leftmargin=*, itemsep=0pt, topsep=0pt]
    \item \textbf{Wrong Commands}: Models generate commands that result in errors or stop the investigation; e.g., using incorrect tool syntax or parameters that cause execution errors.
    
    \item \textbf{Infinite Loop}: Models get stuck repeating the same or very similar commands without making meaningful progress toward solving the problem.
    
    \item \textbf{Hallucination}: Models fabricate answers \cite{rawte2023survey, huang2023survey} or create fictional simulations without any actual evidence.

    \item \textbf{Shallow Investigation}: Models tend to oversimplify tasks \cite{allaway2024exceptions} by relying on obvious keywords rather than performing a thorough analysis. 
    Additionally, the models often fail to explore alternative analytical paths that could lead to more accurate or comprehensive insights.


    \item \textbf{Question Dependency}: When the model fails to answer the initial question, all subsequent dependent analyses go wrong, such as failing to extract a malicious PDF from network capture restricts
    further analysis on that PDF.
\end{itemize}

Table \ref{tab:llmfailure} summarizes the frequency of the failure modes and their top three underlying causes across models. 

\noindent \textit{\ul{Round Exceed:}} 
This issue primarily affects Llama models, which often become stuck in infinite loops. Llama-3.1-8B fails this way in 62.9\% of cases, while larger variants improve (Llama-3.1-405B: 22.7\%). Llama-3.1-8B repeatedly attempts the same question and re-runs identical commands without progressing. In contrast, the other eight models rarely exhibit this behavior, indicating more effective tool usage and interaction.


\noindent \textit{\ul{Give Up:}} 
Gemini-1.5-pro shows the highest question abandonment rate (26.2\%), followed by GPT-4o (25.8\%) and GPT-4o-mini (24.9\%), typically quitting after 4--5 failed attempts due to repetitive incorrect commands or shallow investigation. In contrast, Claude-3.5-Sonnet is more persistent (7.4\% abandonment). Newer models, such as Claude-4.5-Sonnet and GPT-5, rarely give up and instead take additional steps and explore alternative strategies to solve a question (see Appendix~\ref{sec:steps_needed}).


\noindent \textit{\ul{Wrong Answer:}} 
o3-mini has the highest wrong-answer rate (58.1\%), followed by GPT-4o-mini (46.3\%) and DeepSeek-Reasoner (43.2\%), mainly due to question-dependency failures and shallow analysis. GPT-4o, Claude-3.5-Sonnet, GPT-5, and Claude-4.5-Sonnet perform better (19.2\%, 23.6\%, 14.94\%, 11.47\%) through deeper reasoning. For instance, weaker models pick the IP with most packets, while stronger models analyze unique port-scanning patterns for accurate detection.

\begin{table}[h]
\centering
\fontsize{6.5}{7}\selectfont  
\setlength{\arrayrulewidth}{0.15pt}  
\begin{adjustbox}{width=\columnwidth,center}
\begin{tabular}{|p{1.5cm}|p{0.6cm}|p{0.9cm}|p{0.8cm}|p{0.7cm}|p{2.6cm}|}
\hline
\textbf{Model} & \textbf{Give Up} & \textbf{Wrong Answer} & \textbf{Round Exceed} & \textbf{Total failure} & \textbf{Top 3 Failing Reasons (ordered)} \\ \hline
Llama-3.1-8B & 8.3\% & 23.1\% & 62.9\% & 94.3\% & Infinite Loop, Wrong Commands, Hallucination \\ \hline
Llama-3.1-70B & 9.2\% & 40.17\% & 34.1\% & 83.47\% & Infinite Loop, Wrong Commands, Hallucination \\ \hline
Llama-3.1-405B & 11.4\% & 42.62\% & 22.7\% & 76.72\% & Infinite Loop, Wrong Commands, Question Dependency \\ \hline
GPT-4o-mini & 24.9\% & 46.3\% & 0\% & 71.2\% & Question Dependency, Wrong Commands, Shallow Investigation \\ \hline
Gemini-1.5-Pro & 26.2\% & 35.8\% & 0\% & 62.0\% & Question Dependency, Shallow Investigation, Wrong Commands \\ \hline
o3-mini & 0.9\% & 58.1\% & 0.4\% & 59.4\% & Shallow Investigation, Question Dependency, Hallucination \\ \hline
DeepSeek-Reasoner & 6.6\% & 43.2\% & 1.7\% & 51.5\% & Question Dependency, Shallow Investigation, Wrong Commands \\ \hline
GPT-4o & 25.8\% & 19.2\% & 0.4\% & 45.4\% & Question Dependency, Shallow Investigation, Wrong Commands \\ \hline
Claude-3.5-Sonnet & 7.4\% & 23.6\% & 0\% & 31.0\% & Question Dependency, Shallow Investigation, Wrong Commands \\ \hline
GPT-5 & 0.44\% & 14.94\% & 3.92\% & 19.3\% & Question Dependency, Wrong Commands \\ \hline
Claude-4.5 & 1.61\% & 11.47\% & 5.22\% & 18.3\% & Question Dependency, Wrong Commands \\ \hline
\end{tabular}
\end{adjustbox}
\caption{Main Failing Reasons for LLMs. Total failure = sum of failures.}
         \vspace{-0.2in}
\label{tab:llmfailure}
\end{table}

\subsubsection{Consistency in LLM's Performance (RQ4)}
\begin{figure}[htp]
    \vspace{-0.1in}

    \centering
    \includegraphics[width=\columnwidth]{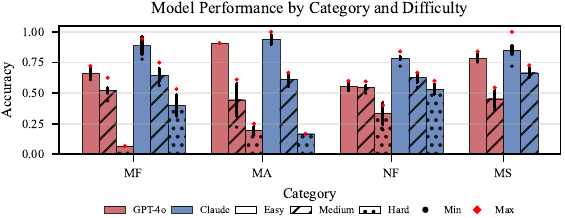}
    \vspace{-0.2in}
    \caption{Measuring Consistency in LLM Performance.}
    \vspace{-0.1in}
    \label{fig:consistency}
\end{figure}

Despite configuring LLMs with deterministic settings with a temperature value of zero and a fixed seed value, output variability remains as a well-documented phenomenon. Muennighoff et al. \cite{muennighoff2025s1} show that even with identical seeds and greedy sampling, LLM responses can vary significantly. Similarly, AutoPenBench \cite{gioacchini2024autopenbench} highlights a similar phenomenon in the context of agent-based evaluations. Motivated by these findings, we conduct an additional analysis to assess the consistency of agents on SIA tasks. Specifically, we evaluate the top-performing models: GPT-4o and Claude-3.5-Sonnet by executing each task three times under identical conditions. The results are summarized in Fig. \ref{fig:consistency}, which 
captures performance variability across multiple runs.

Overall, both models show relatively low variance across most task categories, indicating stable behavior in repeated task executions. However, a closer inspection reveals that Claude-3.5-Sonnet demonstrates higher consistency than GPT-4o, as reflected by smaller error bars across the majority of task categories. Notably, as task difficulty increases from easy to hard, GPT-4o exhibits larger performance fluctuations, particularly in medium and hard tasks. 
In our scenarios, we observe that the LLMs do not consistently use the same ways/commands to solve similar questions. Even a single incorrect command can lead to compounding errors, significantly altering the final outcome. 


\subsection{LLM's Performance on Live SIA Tasks (RQ5)} \label{subsec:RQ5}

RQ5 serves two primary objectives: (i) to address concerns about potential data contamination by evaluating models on challenges published after their respective knowledge cutoffs, and (ii) to demonstrate that benchmark performance translates effectively to live SIA tasks (i.e., tasks still active on the corresponding platforms). We evaluate the top three models from our top four: GPT-5, Claude-3.5-Sonnet and GPT-4o from \oursolution on SIA tasks using active Network Forensics challenges from LetsDefend \cite{LetsDefend}, CyberDefenders \cite{CyberDefenders} and Malware Traffic Analysis \cite{malware_traffic} platforms. Claude-4.5-Sonnet is excluded from evaluation because its knowledge cutoff (July 2025) limits the availability of related live challenges with verified ground-truth answers.



We select seven live challenges (two easy, four medium, and one hard, total 68 questions) published after the knowledge cutoff date of Claude-3.5-Sonnet (April 2024) \cite{anthropic} and GPT-4o (September 30, 2023)\cite{OpenAI}). For GPT-5 (knowledge cutoff - September 2024), we select four additional challenges (three Easy, one with unspecified difficulty; total 20 questions) released after September 2024. All problems were presented in their original format, consistent with the style used in our SIA dataset, without modifications to scenario descriptions or questions. All required artifacts were downloaded locally for experimentation.
Table \ref{tab:active_challenges} summarizes performance on the live tasks across different difficulty levels in two sections: post-April-2024 (7 scenarios) and post-September-2024 (4 scenarios). Among the two models evaluated on the post-April-2024 set, Claude-3.5-Sonnet\footnote{Claude-3.5-Sonnet has been deprecated as of October 2025. So, we were unable to evaluate with new tasks.} significantly outperforms GPT-4o, achieving 60.2\% overall accuracy compared to GPT-4o’s 36.7\%. 
This gap is consistent across Easy and Medium scenarios (e.g., 6/9 vs. 3/9 in Compromised Network Printer, 8/14 vs. 4/14 in Malicious WordPress Plugin). Both models exhibit degraded performance in the Hard scenario (Compromised ICS Device), where each solves 4 of 10 questions, reflecting how increasing complexity reduces model effectiveness. For the post-September-2024 challenges, GPT-5 demonstrates strong performance, correctly answering nearly all questions (e.g., 5/5 in VoIP and 6/6 in Malware Traffic) compared to GPT-4o, highlighting the improved reasoning capabilities of the latest releases.

The results also align with findings from the SIA benchmark: Claude-3.5-Sonnet consistently outperforms GPT-4o and struggle to solve full scenarios. This consistency supports the validity of both our SIA Task dataset and evaluation methodology, indicating minimal influence from potential data contamination such as inclusion of write-ups in training data. 

\begin{table}[]
\fontsize{12}{15}\selectfont
\setlength{\arrayrulewidth}{0.25pt}
\setlength{\tabcolsep}{6pt}
\renewcommand{\arraystretch}{1.35}
\centering
\begin{adjustbox}{width=\columnwidth,center}
\begin{tabular}{|l|c|c|c|c|c|}
\hline
\textbf{Scenario Name} & \textbf{Platform} & \textbf{Difficulty} & \textbf{\begin{tabular}[c]{@{}c@{}}Claude-3.5\\ -Sonnet\\ SQ/TQ\end{tabular}} & \textbf{\begin{tabular}[c]{@{}c@{}}GPT-4o\\ SQ/TQ\end{tabular}} & \textbf{\begin{tabular}[c]{@{}c@{}}GPT-5\\ SQ/TQ\end{tabular}} \\ \hline
\multicolumn{6}{|c|}{\textbf{After training cutoff date of April 2024}} \\ \hline
Compromised Network Printer & LetsDefend & Easy & 6/9 & 3/9 & - \\ \hline
PHP-CGI (CVE-2024-4577) & LetsDefend & Easy & 8/9 & 5/9 & - \\ \hline
Brute Force Attacks & LetsDefend & Medium & 3/8 & 2/8 & - \\ \hline
Compromised Chat Server & LetsDefend & Medium & 6/10 & 3/10 & - \\ \hline
Malicious Traffic Analysis & LetsDefend & Medium & 6/8 & 4/8 & - \\ \hline
Malicious WordPress Plugin & LetsDefend & Medium & 8/14 & 4/14 & - \\ \hline
Compromised ICS Device & LetsDefend & Hard & 4/10 & 4/10 & - \\ \hline
\textbf{Accuracy over 7 tasks} & & & \textbf{41/68 (60.3\%)} & \textbf{25/68 (36.8\%)} & \textbf{-} \\ \hline
\multicolumn{6}{|c|}{\textbf{After training cutoff date of September 2024}} \\ \hline
VoIP & LetsDefend & Easy & - & 5/5 & 5/5 \\ \hline
Lockdown Lab & CyberDefenders & Easy & - & 3/4 & 4/4 \\ \hline
XLMRat Lab & CyberDefenders & Easy & - & 2/5 & 2/5 \\ \hline
Malware Traffic & Malware Traffic Analysis & - & - & 5/6 & 6/6 \\ \hline
\textbf{Accuracy over 4 tasks} & & & \textbf{-} & \textbf{15/20 (75.0\%)} & \textbf{17/20 (85.0\%)} \\ \hline
\end{tabular}
\end{adjustbox}
\footnotesize{SQ = Solved Questions, TQ = Total Questions}
\caption{Performance on tasks after training cutoff date}
\label{tab:active_challenges}
\end{table}

\subsection{Ablation Study on \oursolution Agent (RQ6)} \label{subsec:ablationstudy}
We also investigate how our agent design such as multi-state, ReAct reasoning, and summarization, influences the LLM's performance evaluation. We follow the standard process of ablation study, where performance is measured with the presence and absence of the component. In this section, we conduct experiments on Claude-3.5-Sonnet, GPT-4o, and one latest reasoning model, GPT-5.



\subhead{Multi-State vs. Single-State Workflow} We evaluate \oursolution agent with single-state and multi-state workflow (as discussed in Sec. \ref{sec:agent}). 
Table \ref{tab:single_state_vs_multi_state} demonstrates that our agent with multi-state consistently outperforms the agent with single-state for GPT-5, Claude-3.5-Sonnet and GPT-4o. 
The result infers that multi-state workflow reduces context complexity, enabling the agent's focused reasoning on the current question resulting better solving percentage.

\begin{table}[htbp]
\centering
\fontsize{3.5}{4}\selectfont  
\setlength{\arrayrulewidth}{0.15pt}  
\begin{adjustbox}{width=\columnwidth,center}
\begin{tabular}{|l|l|c|c|c|c|}
\hline
\textbf{Workflow} & \textbf{Model} & \textbf{\begin{tabular}[c]{@{}c@{}}Memory\\Forensics\end{tabular}} & \textbf{\begin{tabular}[c]{@{}c@{}}Malware\\Analysis\end{tabular}} & \textbf{\begin{tabular}[c]{@{}c@{}}Network\\Forensics\end{tabular}} & \textbf{Miscellaneous} \\ \hline
\multirow{3}{*}{\textbf{\begin{tabular}[c]{@{}l@{}}Multi-State\\(MS)\end{tabular}}} & Claude-3.5-Sonnet & 70.0\% & 68.93\% & 68.63\% & \textbf{80.68\%} \\ \cline{2-6}
& GPT-4o & 57.0\% & 58.93\% & 50.6\% & 72.7\% \\ \cline{2-6}
& GPT-5 & \textbf{76.31\%} & \textbf{75.71\%} & \textbf{88.40\%} & 80.49\% \\ \hline
\multirow{3}{*}{\textbf{\begin{tabular}[c]{@{}l@{}}Single-State\\(SS)\end{tabular}}} & Claude-3.5-Sonnet & 45.9\% & 55.5\% & 52.84\% & 59.12\% \\ \cline{2-6}
& GPT-4o & 41.7\% & 40.3\% & 29.68\% & 48.9\% \\ \cline{2-6}
& GPT-5 & 62.36\% & 61.07\% & 74.69\% & 69.32\% \\ \hline
\end{tabular}
\end{adjustbox}
\caption{LLMs' Performance Comparison for Single State and Multi State Workflow.}
         \vspace{-0.1in}
\label{tab:single_state_vs_multi_state}
\end{table}

\subhead{Summarizer vs. No Summarizer} We
evaluate \oursolution agent with the presence and absence of the summarizer component.  
Table \ref{tab:summarizer_effect} shows that the Claude-3.5-Sonnet and GPT-4o models encounter context limitation error (CLE) without the summarizer, while no CLE is encountered when the summarizer module is leveraged by the agent. While calculating the PS, we consider the number of questions the model is able to solve before encountering the CLE, so the partial solving percentage is not zero even if the model encounters CLE at some point. Table \ref{tab:summarizer_effect} demonstrates that the PS of the models improves by 12.52\% to 31.93\% (for Claude-3.5-Sonnet) when the summarizer module is used, and the encounter of CLE is zero. The results show that the summarizer provides key security insights without losing important details from the analysis report, mitigating context limitation and improving overall performance. In contrast, more recent model like GPT-5, due to its substantially large context window, ($400k$ vs $128k$) does not encounter CLE in any task. This larger context allows it to handle entire inputs directly in itself showing better performance without summarizer in some tasks, e.g., Malware Analysis or Miscellaneous tasks involving little visible data (e.g., code snippets and email) without losing information. However, in contrast, areas like network or memory forensics produce large logs where summarization remains useful.

\begin{table}[htbp]
\centering
\resizebox{\columnwidth}{!}{%
\begin{tabular}{|l|l|c|c|c|c|c|c|c|c|}
\hline
\multirow{2}{*}{\textbf{Summarizer}} & \multirow{2}{*}{\textbf{Model}} & \multicolumn{2}{c|}{\textbf{\begin{tabular}[c]{@{}c@{}}Memory\\Forensics\end{tabular}}} & \multicolumn{2}{c|}{\textbf{\begin{tabular}[c]{@{}c@{}}Malware\\Analysis\end{tabular}}} & \multicolumn{2}{c|}{\textbf{\begin{tabular}[c]{@{}c@{}}Network\\Forensics\end{tabular}}} & \multicolumn{2}{c|}{\textbf{Miscellaneous}} \\ \cline{3-10}
 &  & \textbf{CLE} & \textbf{PS} & \textbf{CLE} & \textbf{PS} & \textbf{CLE} & \textbf{PS} & \textbf{CLE} & \textbf{PS} \\ \hline
\multirow{3}{*}{\textbf{With}} & Claude-3.5-Sonnet & 0 & 70.04\% & 0 & 68.93\% & 0 & 68.64\% & 0 & 80.68\% \\ \cline{2-10}
 & GPT-4o & 0 & 57.03\% & 0 & 58.93\% & 0 & 50.60\% & 0 & 72.63\% \\ \cline{2-10}
 & GPT-5 & 0 & \textbf{76.31\%} & 0 & 75.71\% & 0 & \textbf{88.40\%} & 0 & 80.49\% \\ \hline
\multirow{3}{*}{\textbf{Without}} & Claude-3.5-Sonnet & 2 & 54.25\% & 0 & 56.41\% & 3 & 48.23\% & 2 & 48.75\% \\ \cline{2-10}
 & GPT-4o & 5 & 37.1\% & 2 & 43.33\% & 3 & 35.99\% & 1 & 37.92\% \\ \cline{2-10}
 & GPT-5 & 0 & 64.87\% & 0 & \textbf{77.20\%} & 0 & 82.19\% & 0 & \textbf{81.93\%} \\ \hline
\end{tabular}%
}
\footnotesize{CLE = Context Limitation Error, PS = Partially Solved}
\caption{Performance Comparison of LLMs With and Without the Summarizer Module.}
\label{tab:summarizer_effect}
\end{table}



\subhead{ReAct vs. Act-Only} 
We evaluate the \oursolution  agent using both the ReAct (Reason + Action) framework and an Act-Only framework, with results presented in Table \ref{tab:tao_vs_ao}. The findings show that the ReAct framework consistently outperforms the Act-Only approach for Claude-3.5-Sonnet and GPT-4o, indicating improved reasoning capability when the agent is explicitly instructed to think (reason) before acting. In contrast, the performance difference is not significant for GPT-5, as this advanced reasoning model can inherently reason before acting without requiring an explicit ReAct prompt structure. This highlights the continuous improvement of LLMs over time.

\begin{table}[htbp]
\centering
\fontsize{3.5}{4}\selectfont  
\setlength{\arrayrulewidth}{0.15pt}  
\begin{adjustbox}{width=\columnwidth,center}
\begin{tabular}{|l|l|c|c|c|c|}
\hline
\textbf{Method} & \textbf{Model} & \textbf{\begin{tabular}[c]{@{}c@{}}Memory\\Forensics\end{tabular}} & \textbf{\begin{tabular}[c]{@{}c@{}}Malware\\Analysis\end{tabular}} & \textbf{\begin{tabular}[c]{@{}c@{}}Network\\Forensics\end{tabular}} & \textbf{Miscellaneous} \\ \hline
\multirow{3}{*}{\textbf{ReAct}} & Claude-3.5-Sonnet & 70.0\% & 68.9\% & 68.6\% & \textbf{80.68\%} \\ \cline{2-6}
& GPT-4o & 57.0\% & 58.93\% & 50.6\% & 72.63\% \\ \cline{2-6}
& GPT-5 & \textbf{76.31\%} & \textbf{75.71}\% & \textbf{88.40\%} & 80.49\% \\ \hline
\multirow{3}{*}{\textbf{Act-Only}} & Claude-3.5-Sonnet & 46.6\% & 35.25\% & 55.06\% & 56.66\% \\ \cline{2-6}
& GPT-4o & 37.15\% & 30.6\% & 35.99\% & 51.6\% \\ \cline{2-6}
& GPT-5 & 73.97\% & 73.12\% & 81.36\% & 80.49\% \\ \hline
\end{tabular}
\end{adjustbox}
\caption{Comparison Between ReAct vs Act-Only Interaction.}
\label{tab:tao_vs_ao}
\end{table}
These insights can inform design choices and scaffolding strategies when building LLM-based SIA agents. For example, newer and more advanced models may require progressively less scaffolding. Furthermore, guided by the results of the ablation study, future versions of \oursolution will provide modularity, allowing users to select different agent strategies and visualize the impact of various scaffolding approaches.


\subsection{Results on Alert Triaging Task (RQ7)} \label{sec:benign_result}

We evaluate LLMs’ ability to classify alerts as true or false positives using two benchmark datasets: 100 alerts from TII-SRC-23 (50 TP and 50 FP) and 35 alerts from CIC-IDS2017 (5 TP and 30 FP). We conduct this evaluation using \oursolution with two top-performing models (Claude-4.5-Sonnet and GPT-5), one open-weight reasoning model (DeepSeek-Reasoner), and one lightweight model (GPT-4o-mini).


The results in Table \ref{tab:alert_evaluation_combined} demonstrate strong overall performance across both datasets, with notable consistency in model rankings. On the TII-SRC-23 dataset, GPT-5 achieves the highest accuracy at 98\%, demonstrating exceptional performance in identifying both true positives (96\%) and false positives (100\%). Claude-4.5-Sonnet follows closely with 96\% accuracy, achieving perfect true positive identification (100\%) while correctly classifying 46/50 false positive alerts (92\%). On the CIC-IDS2017 dataset, GPT-5 maintains superior performance with 97.1\% accuracy, achieving perfect false positive detection (100\%) and identifying 4/5 true positives (80\%). Claude-4.5-Sonnet demonstrates consistent performance with 88.57\% accuracy, correctly identifying 4/5 true positives (80\%) and 27/30 false positives (90\%).
DeepSeek-Reasoner shows consistent mid-tier performance across both datasets, achieving 89\% accuracy on TII-SRC-23 (47/50 TP, 42/50 FP) and 85.7\% on CIC-IDS2017 (4/5 TP, 26/30 FP). GPT-4o-mini shows the weakest performance with 75\% accuracy on TII-SRC-23 and 51.4\% on CIC-IDS2017, particularly struggling with false positive detection (32/50 and 14/30, respectively), suggesting a bias toward treating alerts as genuine threats.

The superior performance of GPT-5 in achieving near-perfect false positive detection while maintaining high true positive accuracy across both datasets is particularly noteworthy for SOC operations. The consistent performance gap between advanced models and less capable alternatives highlights the importance of model selection for security operations, where the ability to accurately distinguish false positives from true threats directly impacts analyst efficiency and resource allocation.



\begin{table}[h]
\centering
\setlength{\arrayrulewidth}{0.15pt}
\fontsize{3.5}{4}\selectfont
\begin{adjustbox}{width=\columnwidth,center}
\begin{tabular}{|l|c|c|c|c|c|}
\hline
\textbf{Model} & \multicolumn{2}{c|}{\textbf{TP}} & \multicolumn{2}{c|}{\textbf{FP}} & \textbf{Accuracy} \\ \hline
\multicolumn{6}{|c|}{\textbf{TII-SRC-23 Dataset (50 TP and 50 FP)}} \\ \hline
Claude-4.5-Sonnet & 50/50 & 100\% & 46/50 & 92\% & 96\% \\ \hline
GPT-5 & 48/50 & 96\% & 50/50 & 100\% & 98\% \\ \hline
DeepSeek-Reasoner & 47/50 & 94\% & 42/50 & 84\% & 89\% \\ \hline
GPT-4o-mini & 43/50 & 86\% & 32/50 & 64\% & 75\% \\ \hline
\multicolumn{6}{|c|}{\textbf{CIC-IDS2017 Dataset (5 TP and 30 FP)}} \\ \hline
Claude-4.5-Sonnet & 4/5 & 80\% & 27/30 & 90.0\% & 88.57\% \\ \hline
GPT-5 & 4/5 & 80\% & 30/30 & 100\% & 97.1\% \\ \hline
DeepSeek-Reasoner & 4/5 & 80\% & 26/30 & 86.7\% & 85.7\% \\ \hline
GPT-4o-mini & 4/5 & 80\% & 14/30 & 46.67\% & 51.4\% \\ \hline
\end{tabular}
\end{adjustbox}
\caption{Alert Triaging Results across TII-SRC-23 and CIC-IDS2017 Datasets (correct/total).}
\label{tab:alert_evaluation_combined}
\end{table}

\section{Related Works}
\label{sec:relatedwork}
In this section, we review existing works that
building benchmarks and automation tools using LLMs
covering various security application domains,
generating benchmarking datasets, building
agentic workflow, and evaluating LLM capabilities.
We further compare those works with \oursolution in Table \ref{tab:comparison} in Appendix \ref{appendix:explanation}.

\subhead{Application Domains} 
Existing cybersecurity benchmarks target distinct application domains. 
Security knowledge benchmarks like CyberMetric \cite{tihanyi2024cybermetric} and CyberBench \cite{liu2024cyberbench} evaluate foundational cybersecurity concepts through MCQs and NER tasks. Offensive security works \cite{deng2024pentestgpt, gioacchini2024autopenbench, zhang2024cybench, yang2023language} focus on offensive scenarios like exploiting vulnerable machines and solving CTF challenges. Threat intelligence benchmarks like CTIBench~\cite{alam2024ctibench}, AthenaBench~\cite{alam2025athenabench}, and CyberSOCEval~\cite{deason2025cybersoceval} target tasks such as MITRE ATT\&CK mapping, CVSS scoring. CyberSOCEval also addresses analyzing threat reports through MCQ questions. Code analysis works \cite{hu2024degpt, code_analysis, secllmholm} evaluate LLMs on decompiled code translation, obfuscated code explanation or identifying security bugs from code snippets. IRCopilot \cite{lin2025ircopilot} targets incident response tasks having limited domain coverage. The primary focus of \oursolution is the investigative aspect of security incidents, covering major task areas and incorporating both benign and malicious artifacts aligned with real-world SOC scenarios.


\subhead{Dataset Generation} Code analysis and threat intelligence benchmarks \cite{code_analysis, alam2024ctibench, secllmholm, alam2025athenabench, deason2025cybersoceval} are non-agentic, typically focus on static, single-step tasks.
There are other agentic benchmark frameworks \cite{zhang2024cybench, gioacchini2024autopenbench, yang2023language} address multi-step tasks, different offensive tasks, or CTF challenges from multiple domains with varying difficulty. These works often address single objectives (e.g., find a flag), whereas our work, \oursolution, targets uncovering multiple objectives (e.g., who, what, when-based questions). Whereas, automation tools work, such as \cite{deng2024pentestgpt, lin2025ircopilot} have a dataset of limited scale. While \cite{lin2025ircopilot} addresses multi-objective tasks across 12 scenarios and 130 subquestions, their dataset often includes biased (as-is from the platform sources) and redundant questions (e.g., treating the identification of attacker's two IP addresses as separate subtasks). In contrast, our work introduces a more refined and comprehensive benchmark by preprocessing the data to reduce redundancy and bias. We also expand the scope significantly, covering 25 scenarios and 229 questions across multiple key domains, and uniquely include alert triaging tasks with 55 true positive and 80 false positive scenarios-offering a broader and more realistic benchmark dataset for SIA tasks.


\subhead{Agentic Workflow} 
Although automation tools like \cite{deng2024pentestgpt} and \cite{lin2025ircopilot} use LLM APIs, 
they rely 
on human professionals to ask questions to the LLM and execute the action in the environment, which does not necessarily evaluate how good LLMs are at taking actions autonomously. Whereas benchmarks \cite{gioacchini2024autopenbench, liu2024cyberbench, yang2023language} develop automated agents with tool access to automatically perform the tasks. Aligning with these works, for SIA tasks, we develop an agent to address multi-question scenarios and execute the tasks autonomously with specialized capabilities: having environment and tool access for interacting with the environment and analyzing different types of artifacts, context optimization to filter relevant information from verbose tool outputs (e.g., long logs) and reduce the context-limitation problem.

\subhead{Evaluation} Code analysis works \cite{code_analysis, hu2024degpt, secllmholm} evaluate with both malicious and benign code snippets. Other works, they do not consider benign samples. For instance, AutoPenBench and PentestGPT \cite{gioacchini2024autopenbench, deng2024pentestgpt} exclusively evaluate LLM performance on vulnerable machines. Similarly, CTF-based benchmarks such as CyBench~\cite{zhang2024cybench} and IntercodeCTF~\cite{yang2023language} utilize challenges that are intentionally crafted to contain vulnerabilities to evaluate the LLMs on offensive tasks.
IRCopilot \cite{lin2025ircopilot} focuses only on malicious samples for analysis.
Whereas, \oursolution 
incorporates both malicious and benign samples in LLM evaluation for SIA tasks with a better alignment with real-world. 

In contrast to existing works, \oursolution focuses specifically on security incident
analysis tasks with multiple dataset design and agentic
workflow features, with evaluation of both benign and
attack artifacts, addressing a gap not covered by
other security domain benchmarks.

\section{Conclusion}

This paper presented \oursolution, an agentic benchmarking framework for evaluating the capability of LLMs in  performing Security Incident Analysis (SIA). We introduced a benchmark dataset covering diverse SIA scenarios, including alert triage, and developed an automated agent that executes multi-objective investigations, integrates with 11 major LLMs, and supports future extensions.
Our evaluation highlighted both the progress and limitations of current models: while LLMs are advancing toward automating SIA workflows, improvements in reasoning and reliability remain necessary. \oursolution established baseline performance for off-the-shelf models and enabled consistent tracking of advances, helping researchers and practitioners make informed decisions - such as selecting cost-efficient models, implementing effective guardrails, and identifying when adoption of newer models becomes worthwhile.
Looking ahead, we plan to evaluate more complex SIA scenarios and agent architectures, offering modularity for selecting different agent strategies. 
Furthermore, we will continue expanding our dataset with more diverse SIA incidents and alerts. Additionally, we will maintain our leaderboard~\cite{siabench_link} to continuously keep track of the 
progress of LLM in SIA.

\subhead{Acknowledgement}
This material is based
upon work supported by the Natural Sciences and Engineering
Research Council of Canada (NSERC) and Department of
National Defence Canada (DND) under the Discovery Grants
RGPIN-2021-04106 and DGDND-2021-04106.

\bibliographystyle{IEEEtran}
\bibliography{references}

\appendices

\section{Prompts} \label{sec:prompts}

\subhead{Prompt for Paraphrasing the Scenarios} The prompt is used for content paraphrasing. The prompt: \textit{``Paraphrase the following paragraph, keeping the original meaning intact and maintaining approximately the same length. Do not add or remove any key ideas. Use natural, clear language: [Insert the scenario description here]''}

\subhead{Initial Prompt for \oursolution Agent} A snippet of the prompt provided to the \oursolution agent to initiate its execution is shown in Fig. \ref{fig:agent_prompt}.

\begin{figure*}[htp]
    \footnotesize 
    \centering
   \begin{tcolorbox}[colback=gray!10, colframe=white, boxrule=0pt, left=2mm, right=2mm, top=1mm, bottom=1mm]
        \textbf{System Prompt:} You are a highly skilled cybersecurity professional and cyber-defensive analyst. Your goal is to assist the user in solving cyber-defense–related tasks and answering cybersecurity questions. You will receive questions one by one and must use the tools \texttt{think\_tool}, \texttt{spawn\_shell}, \texttt{execute\_command}, \texttt{save\_answer}, \texttt{give\_up}, and \texttt{next\_question} to find the answers. Before taking any action or using any tool, you must always call \texttt{think\_tool} first. The required flow should follow this pattern: \texttt{think\_tool} before \texttt{spawn\_shell}, \texttt{think\_tool} before \texttt{execute\_command}, and \texttt{think\_tool} before every subsequent step. In other words, every action or decision you make must always be preceded by a reasoning step using {think\_tool}......

        \textbf{User Prompt:} The scenario:\texttt{Scenario\_description}. Tools available and usage instruction: \texttt{tools\_available and instruction}. Files available: \texttt{files\_available}. Directory of the files: \texttt{directory}. The question is: \texttt{question\_asked}
    \end{tcolorbox}
    \caption{Initial Prompt for the \oursolution Agent.}
    \label{fig:agent_prompt}
\end{figure*}

\subhead{Summarizer Module Prompt} Fig. \ref{fig:summarizer_prompt} illustrates a snippet of the prompt provided to the summarizer module in our \oursolution agent.

\begin{figure*}[htp]
    \footnotesize 
    \centering
    \begin{tcolorbox}[colback=gray!10, colframe=white, boxrule=0pt, left=2mm, right=2mm, top=1mm, bottom=1mm]
        \textbf{Summarizer Prompt:} An agent is solving a question in the domain of security incident analysis. The agent has been provided with the following scenario: \texttt{Scenario\_description}. The agent has been asked the following question: \texttt{question\_asked}. The agent has produced the following thought: \texttt{thought}, and issued the following tool command: \texttt{executed\_command}. This command produced the following output: \texttt{tool\_output}. Based on this information, your task is to generate a summary capturing the key points from the command output. Do not provide any suggestions or make assumptions about the final result. Only summarize the key findings to assist in answering the question. Note: The command outputs may be large, so the output might be presented in chunks. If so, you will also be provided with the summary of the previous chunks. The summary of the previous chunks is: \texttt{summary}.
        
    \end{tcolorbox}
    \caption{Prompt for the Summarizer Module.}
    \label{fig:summarizer_prompt}
\end{figure*}

\section{\oursolution Agent Implementation} \label{subsec:agent_implementation}
This section details the implementation of our SIABench agent, developed in a Kali Linux virtual machine environment. The agent is built using LangChain \cite{LangChain}, a widely recognized framework for creating LLM-based agents \cite{topsakal2023creating}, and is developed in Python \cite{python}. Previous works, such as SecGPT \cite{wu2024isolategpt}, also use this framework to build agentic system. The framework provides key functionalities, including (i) parsing outputs in a structured format (schema) and (ii) support tool augmentation. Other frameworks could also be used with minimal changes, e.g., LlamaIndex \cite{LlamaIndex}.

\subhead{Incident Data Module}
This module stores all task-related information, including the scenario description, required artifact names, their directories, tool usage instructions, and questions, in a JSON file. The JSON file serves as input to the LLM.

\subhead{Investigation Planning Module}
To obtain structured outputs like thoughts and actions, we utilize ReAct \cite{yao2022react} and LangChain's output parser, minimizing parsing errors and improving interaction efficiency. We use \texttt{thinking\_tool} functionality to generate the explicit reasoning before generating the specific action. This approach enhances LLM problem-solving by combining reasoning and action-based workflows, consistent with previous works such as \cite{zhang2024cybench} and \cite{gioacchini2024autopenbench}. The LLMs evaluated in this work, as detailed in Sec.~\ref{subsec:experimental_setup}, include OpenAI's GPT models~\cite{OpenAI}, Google's Gemini accessed via its API~\cite{Google_AI}, DeepSeek-Reasoner accessed via its API \cite{deepseek}, Claude models accessed via its API \cite{anthropic}, and Meta's Llama models accessed through the Fireworks API~\cite{fireworks}. All models are seamlessly integrated with LangChain \cite{Chat_models}.

\subhead{Action Execution Module and Analysis Reporting Module} As outlined in our methodology, the Action Execution Module is designed to fulfill three primary functionalities, which are implemented through external tools using a technique known as tool augmentation \cite{zhuang2023toolqa}. We designed these external tools using Python so that the agent can have a seamless interaction between it and the analysis environment to perform the tasks efficiently without any human intervention.

For the Opening and Closing Shell functionality, two distinct external tools are implemented. The first, \darkcircle{1} \texttt{spawn\_shell}, opens a new environment to create isolated spaces for scenario analysis. Upon completion, the \darkcircle{2} \texttt{close\_shell} tool safely and efficiently terminates the environment, ensuring optimal resource utilization. The Execute Tool function is enabled by the \darkcircle{3} \texttt{execute\_command} tool, which allows the agent to execute commands, utilize specialized cybersecurity tools, and download additional tools when required. The State Change functionality uses three external tools to modify the agent's state. The \darkcircle{4} \texttt{answer\_question} tool is invoked when the LLM finds a solution, while the \darkcircle{5} \texttt{give\_up} tool is used when progress is no longer feasible. Together, these tools help the LLM reach the ``Solved'' state, reducing computational effort \cite{nyuctf}, avoiding stagnation, and maintaining focus. The Analysis Reporting Module evaluates responses against ground truth using Gemini-2.5-flash \cite{team2024gemini}, with earlier evaluations performed on Gemini-1.5-flash before its deprecation. It tracks progress and identifies remaining questions. If questions remain, the \darkcircle{6} \texttt{next\_question} tool retrieves the next question from the Incident Data Module, transitioning the agent back to the ``Incident Analysis'' state to resume the analysis process.

These tools enable the agent to emulate a human expert's workflow, interact effectively with its environment, and adaptively solve multi-question scenarios.

\subhead{Summarizer Module}
This module prevents the agent from being overloaded with excess information. It uses the same LLM model as the ``Investigation Planning Module'' to process command outputs by analyzing the agent's thoughts and actions. A GPT-2 tokenizer \cite{tokenizer} segments the output into fixed-size chunks, which are gradually combined to create a comprehensive summary of the analysis.

\section{Discussion} \label{sec:discussion}

\subhead{Model Adaptability and Integration Challenges} \label{sec:newmodel_adaption}Our evaluation framework is designed to be adaptable, supporting various LLMs compatible with LangChain/LangGraph frameworks to enable seamless integration and testing. After developing our SIA agent, we integrate models such as Llama-3.1-8B, Llama-3.1-70B, Llama-3.1-405B, Claude-4.5-Sonnet, Claude-3.5-Sonnet, Gemini-1.5-pro, DeepSeek-Reasoner, GPT-4o-mini, GPT-4o, o3-mini, GPT-5.
Although most models are instruction-tuned \cite{shi2024instruction} and generally adhere to prompts, LLM agents occasionally fail to follow instructions, especially the smaller models like Llama 8B, 70B variants, leading to errors or incomplete task execution \cite{follow_instructions}. 
However, to maintain consistency, in this research, all models are evaluated using the same prompts and agent structure (zero-shot prompting with tool usage instructions and consistent prompt formatting), and we rerun any incomplete runs, if any. Differences in the model's capability and guardrails occasionally poses challenges that may lead to some minor adjustments. For example, by observing that o3-mini struggles with function calling we adjust our prompts to improve the consistency of o3-mini \cite{o3_mini_forum_post}. Also, since GPT-5 flagged our defensive cyber questions as a violation of policy, we apply jail breaking technique to bypass the guardrail to run all tasks, following similar approach of existing works \cite{deng2024pentestgpt}. This also demonstrate the brittleness of the guardrails of modern LLMs.  


\subhead{Mitigating LLM Data Contamination Risk} 
There is a possibility that the evaluated LLMs have seen the write-ups of the \oursolution scenarios in their training data. 
To mitigate such data contamination risk, we adopt three strategies. First, we pre-process the data to minimize triggers that might cause the LLM to retrieve memorized content \cite{kerr2025icarus}, as discussed in section \ref{sec:dataset}. Second, we follow vendor guidance (OpenAI, Microsoft, Anthropic) \cite{anthropic, OpenAI} indicating that model knowledge is limited to its training cutoff unless augmented through tools (e.g., RAG, web access) or explicit fine-tuning. Accordingly, we evaluate models using tasks released after their respective training cutoff dates, as discussed in Section \ref{sec: evaluation_nn}.
Third, we assess whether the LLMs underwent targeted training by querying them by providing scenario name, platform name and raw questions (without the answers) for detailed information about specific scenarios. These conventions have been used in previous works such as PentestGPT, Cybench and IRCopilot \cite{deng2024pentestgpt, zhang2024cybench, lin2025ircopilot}.

\subhead{Guided Nature of Investigative Question}
As outlined in the paper, we evaluate LLMs based on competencies expected of junior-level analysts. Therefore, we selected a diverse SIA task suite comprising Network and Memory Forensics, Malware Analysis, Phishing Investigation, and Log Analysis grounded in training materials used to prepare junior analysts for 5-30 years, ensuring real-world relevance. The underlying assumption \cite{radutest, kerstenfield} is that junior analysts are not required to perform deep investigations; rather, they must demonstrate the ability to follow investigative processes in a guided manner, enabling them to assist senior analysts when tasked with specific questions. While we have taken steps to mitigate bias, some inherent bias persists due to the guided nature of the investigative tasks. Consequently, this approach does not fully capture the unguided, exploratory, tree-like structure characteristic of expert-level SIA tasks, which fall outside the scope of this paper.

\subhead{Security Tool Selection} 
As discussed in Section \ref{sec:dataset}, we recommend some security tools to the LLMs to analyze SIA tasks following blue team training platform standards \cite{CyberDefenders, BTLO, tryhackme}. Since the tool selection depends on context and rarely has a single “right” choice, we evaluate task success using any suitable tool rather than insisting on a specific one, which better reflects practice.
All security tools run inside a controlled Kali Linux VM. The preinstalled packages cover common utilities. For most cases the latest version of the recommended tools is installed, as we do not mention any specific version numbers. LLMs also install the latest version by using default installation commands.


\subhead{Limitation of Using Graphical User Interfaces (GUIs)-based Tools} We only considered command line tools for this work. Incorporating GUI capabilities in future designs could significantly enhance automation and broaden applicability.

\section{Average Steps Needed by LLM} \label{sec:steps_needed} 

Table \ref{tab:steps_tokens} presents the average number of steps taken by top-performing LLMs to solve all the SIA tasks in each difficulty level.

\begin{table}[h]
\vspace{-0.1in}
\centering
\scriptsize 
\setlength{\tabcolsep}{3pt} 
\renewcommand{\arraystretch}{1.1} 
\begin{tabular*}{\columnwidth}{@{\extracolsep{\fill}} llccccc @{}}
\toprule
\textbf{Model}      & \textbf{Level} & \makecell{\textbf{Average} \\ \textbf{Pass} \\ \textbf{Steps}} & \makecell{\textbf{Fail} \\ \textbf{Wrong} \\ \textbf{Steps}} & \makecell{\textbf{Fail} \\ \textbf{Give Up} \\ \textbf{Steps}} \\ \midrule
DeepSeek-Reasoner   & Easy           & 3.480                     & 3.820                               & 3.750                         \\
                    & Medium         & 3.000                     & 3.530                               & 5.500                         \\
                    & Hard           & 3.790                     & 5.500                               & 5.000                         \\ \midrule
GPT-4o             & Easy           & 3.754                     & 3.462                               & 3.000                         \\
                    & Medium         & 3.909                     & 4.000                               & 4.400                         \\
                    & Hard           & 3.444                     & 3.667                               & 5.833                         \\ \midrule
Claude-3.5-Sonnet          & Easy           & 3.420                     & 3.830                               & 5.000                         \\
                    & Medium         & 3.780                     & 5.500                               & 6.000                         \\
                    & Hard           & 4.250                     & 6.090                               & 4.500                         \\ \midrule
GPT-5              & Easy           & 4.115                     & 5.600                               & 4.500                         \\
                    & Medium         & 4.049                     & 4.278                               & 6.000                         \\
                    & Hard           & 5.333                     & 4.357                               & 5.000                         \\ \midrule
Claude-4.5          & Easy           & 4.500                     & 7.333                               & 0.000                         \\
                    & Medium         & 4.494                     & 6.833                               & 2.667                         \\
                    & Hard           & 4.350                     & 7.000                               & 4.667                         \\ \bottomrule

\end{tabular*}
\footnotesize{Fail-Wrong Steps = Avg. Steps for Wrong Answers; Fail-Give Steps = Avg. Steps for Giving Up.}
\caption{Average Steps Calculation for Different Levels of Scenarios.}
\label{tab:steps_tokens}
\end{table}

\section{Comparison with Other Works}
\label{appendix:explanation}

\begin{table*}
\scriptsize
\centering 
\resizebox{\textwidth}{!}{%
\begin{tabular}{|l|c|c|c|c|c|c|c|c|c|c|c|c|c|c|c|c|} 
\hline 
\multirow{2}{*}{\parbox[t][2cm][c]{2cm}{\centering 
\textbf{Works}}} & 
\multicolumn{5}{c|}{\textbf{Application Domain}}  & 
\multicolumn{5}{c|}{\textbf{Dataset Generation}} & 
\multicolumn{4}{c|}{\textbf{Agentic Workflow}} &
\multicolumn{2}{c|}{\textbf{Evaluation}} \\ \cline{2-17}

&
\rotatebox{90}{\makecell{Security \\ Incidents}} & 
\rotatebox{90}{\makecell{Offensive \\ Security}} & 
\rotatebox{90}{\makecell{Threat \\ Intelligence /\\ Analysis}} & 
\rotatebox{90}{\makecell{Code \\ Analysis}} &
\rotatebox{90}{\makecell{Security \\ Knowledge}} &
\rotatebox{90}{\makecell{Data \\ Refinement}} &
\rotatebox{90}{\makecell{Increasing \\ Task \\ Difficulty}} &
\rotatebox{90}{\makecell{Multiple \\ Domain \\ Coverage}} &
\rotatebox{90}{\makecell{Multi \\ Objectives}} &
\rotatebox{90}{\makecell{Multi \\ Steps}} &
\rotatebox{90}{\makecell{LLM \\ Agent}} &
\rotatebox{90}{\makecell{Environment \\ Access}} &
\rotatebox{90}{\makecell{Tool \\ Access}} &
\rotatebox{90}{\makecell{Context \\ Optimize}} &
\rotatebox{90}{\makecell{Actual Benign \\ (Non-Vuln)}} &
\rotatebox{90}{\makecell{Actual \\ Malicious \\ (Vuln)}} \\
\hline

\multicolumn{17}{|c|}{\textbf{Benchmarking Frameworks}} \\
\hline

\textbf{\oursolution} &
{\scalebox{0.9}{\ding{108}}} &
\raisebox{-0.35ex}{\scalebox{1.75}{$\circ$}} &
\raisebox{-0.35ex}{\scalebox{1.75}{$\circ$}} & 
\raisebox{-0.35ex}{\scalebox{1.75}{$\circ$}} & 
\raisebox{-0.35ex}{\scalebox{1.75}{$\circ$}} &
{\scalebox{0.9}{\ding{108}}} &
{\scalebox{0.9}{\ding{108}}} &
{\scalebox{0.9}{\ding{108}}} &
{\scalebox{0.9}{\ding{108}}} &
{\scalebox{0.9}{\ding{108}}} &
{\scalebox{0.9}{\ding{108}}} & 
{\scalebox{0.9}{\ding{108}}} &
{\scalebox{0.9}{\ding{108}}} & 
{\scalebox{0.9}{\ding{108}}} &
{\scalebox{0.9}{\ding{108}}} &
{\scalebox{0.9}{\ding{108}}} \\
\hline

{AutoPenBench}~\cite{gioacchini2024autopenbench} &
\raisebox{-0.35ex}{\scalebox{1.75}{$\circ$}} &
{\scalebox{0.9}{\ding{108}}} &
\raisebox{-0.35ex}{\scalebox{1.75}{$\circ$}} & 
\raisebox{-0.35ex}{\scalebox{1.75}{$\circ$}} & 
\raisebox{-0.35ex}{\scalebox{1.75}{$\circ$}} &
\raisebox{-0.35ex}{\scalebox{1.75}{$\circ$}} &
{\scalebox{0.9}{\ding{108}}} &
{\scalebox{0.9}{\ding{108}}} &
\raisebox{-0.35ex}{\scalebox{1.75}{$\circ$}} &
{\scalebox{0.9}{\ding{108}}} &
{\scalebox{0.9}{\ding{108}}} &
{\scalebox{0.9}{\ding{108}}} &
{\scalebox{0.9}{\ding{108}}} &
{\scalebox{0.9}{\ding{108}}} &
\raisebox{-0.35ex}{\scalebox{1.75}{$\circ$}} &
{\scalebox{0.9}{\ding{108}}} \\
\hline

{InterCodeCTF}~\cite{yang2023language} &
\raisebox{-0.35ex}{\scalebox{1.75}{$\circ$}} &
{\scalebox{0.9}{\ding{108}}} &
\raisebox{-0.35ex}{\scalebox{1.75}{$\circ$}} & 
\raisebox{-0.35ex}{\scalebox{1.75}{$\circ$}} & 
\raisebox{-0.35ex}{\scalebox{1.75}{$\circ$}} &
\raisebox{-0.35ex}{\scalebox{1.75}{$\circ$}} &
{\scalebox{0.9}{\ding{108}}} &
{\scalebox{0.9}{\ding{108}}} &
\raisebox{-0.35ex}{\scalebox{1.75}{$\circ$}} &
{\scalebox{0.9}{\ding{108}}} &
{\scalebox{0.9}{\ding{108}}} &
{\scalebox{0.9}{\ding{108}}} &
{\scalebox{0.9}{\ding{108}}} &
\raisebox{-0.35ex}{\scalebox{1.75}{$\circ$}} &
\raisebox{-0.35ex}{\scalebox{1.75}{$\circ$}} &
{\scalebox{0.9}{\ding{108}}} \\
\hline

{Cybench}~\cite{zhang2024cybench} &
\raisebox{-0.35ex}{\scalebox{1.75}{$\circ$}} &
{\scalebox{0.9}{\ding{108}}} &
\raisebox{-0.35ex}{\scalebox{1.75}{$\circ$}} & 
\raisebox{-0.35ex}{\scalebox{1.75}{$\circ$}} &
\raisebox{-0.35ex}{\scalebox{1.75}{$\circ$}} &
\raisebox{-0.35ex}{\scalebox{1.75}{$\circ$}} &
{\scalebox{0.9}{\ding{108}}} &
{\scalebox{0.9}{\ding{108}}} &
\raisebox{-0.35ex}{\scalebox{1.75}{$\circ$}} &
{\scalebox{0.9}{\ding{108}}} &
{\scalebox{0.9}{\ding{108}}} &
{\scalebox{0.9}{\ding{108}}} &
{\scalebox{0.9}{\ding{108}}} &
\raisebox{-0.35ex}{\scalebox{1.75}{$\circ$}} &
\raisebox{-0.35ex}{\scalebox{1.75}{$\circ$}} &
{\scalebox{0.9}{\ding{108}}} \\
\hline

{CyberMetric}~\cite{tihanyi2024cybermetric}
&
\raisebox{-0.35ex}{\scalebox{1.75}{$\circ$}} &
\raisebox{-0.35ex}{\scalebox{1.75}{$\circ$}} &
\raisebox{-0.35ex}{\scalebox{1.75}{$\circ$}} &
\raisebox{-0.35ex}{\scalebox{1.75}{$\circ$}} &
{\scalebox{0.9}{\ding{108}}} &
{\scalebox{0.9}{\ding{108}}} &
\raisebox{-0.35ex}{\scalebox{1.75}{$\circ$}} &
\raisebox{-0.35ex}{\scalebox{1.75}{$\circ$}} &
\raisebox{-0.35ex}{\scalebox{1.75}{$\circ$}} &
\raisebox{-0.35ex}{\scalebox{1.75}{$\circ$}} &
\raisebox{-0.35ex}{\scalebox{1.75}{$\circ$}} & 
\raisebox{-0.35ex}{\scalebox{1.75}{$\circ$}} &
\raisebox{-0.35ex}{\scalebox{1.75}{$\circ$}} &
\raisebox{-0.35ex}{\scalebox{1.75}{$\circ$}} &
\raisebox{-0.35ex}{\scalebox{1.75}{$\circ$}} &
\raisebox{-0.35ex}{\scalebox{1.75}{$\circ$}} \\
\hline

{CyberBench}~\cite{liu2024cyberbench} &
\raisebox{-0.35ex}{\scalebox{1.75}{$\circ$}} &
\raisebox{-0.35ex}{\scalebox{1.75}{$\circ$}} &
\raisebox{-0.35ex}{\scalebox{1.75}{$\circ$}} &
\raisebox{-0.35ex}{\scalebox{1.75}{$\circ$}} &
{\scalebox{0.9}{\ding{108}}} &
\raisebox{-0.35ex}{\scalebox{1.75}{$\circ$}} &
\raisebox{-0.35ex}{\scalebox{1.75}{$\circ$}} &
\raisebox{-0.35ex}{\scalebox{1.75}{$\circ$}} &
\raisebox{-0.35ex}{\scalebox{1.75}{$\circ$}} &
\raisebox{-0.35ex}{\scalebox{1.75}{$\circ$}} &
\raisebox{-0.35ex}{\scalebox{1.75}{$\circ$}} & 
\raisebox{-0.35ex}{\scalebox{1.75}{$\circ$}} &
\raisebox{-0.35ex}{\scalebox{1.75}{$\circ$}} &
\raisebox{-0.35ex}{\scalebox{1.75}{$\circ$}} &
\raisebox{-0.35ex}{\scalebox{1.75}{$\circ$}} &
\raisebox{-0.35ex}{\scalebox{1.75}{$\circ$}} \\
\hline

{Code Analysis}~\cite{code_analysis} &
\raisebox{-0.35ex}{\scalebox{1.75}{$\circ$}} &
\raisebox{-0.35ex}{\scalebox{1.75}{$\circ$}} &
\raisebox{-0.35ex}{\scalebox{1.75}{$\circ$}} & 
{\scalebox{0.9}{\ding{108}}} & 
\raisebox{-0.35ex}{\scalebox{1.75}{$\circ$}} &
{\scalebox{0.9}{\ding{108}}} &
{\scalebox{0.9}{\ding{108}}} &
\raisebox{-0.35ex}{\scalebox{1.75}{$\circ$}} &
\raisebox{-0.35ex}{\scalebox{1.75}{$\circ$}} &
\raisebox{-0.35ex}{\scalebox{1.75}{$\circ$}} &
\raisebox{-0.35ex}{\scalebox{1.75}{$\circ$}} & 
\raisebox{-0.35ex}{\scalebox{1.75}{$\circ$}} &
\raisebox{-0.35ex}{\scalebox{1.75}{$\circ$}} & 
\raisebox{-0.35ex}{\scalebox{1.75}{$\circ$}} &
{\scalebox{0.9}{\ding{108}}} &
{\scalebox{0.9}{\ding{108}}} \\
\hline

{CTIBench}~\cite{alam2024ctibench} & 
\raisebox{-0.35ex}{\scalebox{1.75}{$\circ$}} &
\raisebox{-0.35ex}{\scalebox{1.75}{$\circ$}} &
{\scalebox{0.9}{\ding{108}}} & 
\raisebox{-0.35ex}{\scalebox{1.75}{$\circ$}} & 
\raisebox{-0.35ex}{\scalebox{1.75}{$\circ$}} &
{\scalebox{0.9}{\ding{108}}} &
{\scalebox{0.9}{\ding{108}}} &
\raisebox{-0.35ex}{\scalebox{1.75}{$\circ$}} &
\raisebox{-0.35ex}{\scalebox{1.75}{$\circ$}} &
\raisebox{-0.35ex}{\scalebox{1.75}{$\circ$}} &
\raisebox{-0.35ex}{\scalebox{1.75}{$\circ$}} & 
\raisebox{-0.35ex}{\scalebox{1.75}{$\circ$}} & 
\raisebox{-0.35ex}{\scalebox{1.75}{$\circ$}} &
\raisebox{-0.35ex}{\scalebox{1.75}{$\circ$}} &
\raisebox{-0.35ex}{\scalebox{1.75}{$\circ$}} &
\raisebox{-0.35ex}{\scalebox{1.75}{$\circ$}} \\
\hline

{AthenaBench}~\cite{alam2025athenabench} & 
\raisebox{-0.35ex}{\scalebox{1.75}{$\circ$}} &
\raisebox{-0.35ex}{\scalebox{1.75}{$\circ$}} &
{\scalebox{0.9}{\ding{108}}} & 
\raisebox{-0.35ex}{\scalebox{1.75}{$\circ$}} & 
\raisebox{-0.35ex}{\scalebox{1.75}{$\circ$}} &
{\scalebox{0.9}{\ding{108}}} &
\raisebox{-0.35ex}{\scalebox{1.75}{$\circ$}} & 
{\scalebox{0.9}{\ding{108}}} &
\raisebox{-0.35ex}{\scalebox{1.75}{$\circ$}} &
\raisebox{-0.35ex}{\scalebox{1.75}{$\circ$}} &
\raisebox{-0.35ex}{\scalebox{1.75}{$\circ$}} & 
\raisebox{-0.35ex}{\scalebox{1.75}{$\circ$}} & 
\raisebox{-0.35ex}{\scalebox{1.75}{$\circ$}} &
\raisebox{-0.35ex}{\scalebox{1.75}{$\circ$}} &
\raisebox{-0.35ex}{\scalebox{1.75}{$\circ$}} &
{\scalebox{0.9}{\ding{108}}}  \\
\hline

{CyberSOCEval}~\cite{deason2025cybersoceval} & 
\raisebox{-0.35ex}{\scalebox{1.75}{$\circ$}} &
\raisebox{-0.35ex}{\scalebox{1.75}{$\circ$}} &
{\scalebox{0.9}{\ding{108}}} & 
\raisebox{-0.35ex}{\scalebox{1.75}{$\circ$}} & 
\raisebox{-0.35ex}{\scalebox{1.75}{$\circ$}} &
{\scalebox{0.9}{\ding{108}}} &
{\scalebox{0.9}{\ding{108}}} & 
{\scalebox{0.9}{\ding{108}}} &
\raisebox{-0.35ex}{\scalebox{1.75}{$\circ$}} &
\raisebox{-0.35ex}{\scalebox{1.75}{$\circ$}} &
\raisebox{-0.35ex}{\scalebox{1.75}{$\circ$}} & 
\raisebox{-0.35ex}{\scalebox{1.75}{$\circ$}} & 
\raisebox{-0.35ex}{\scalebox{1.75}{$\circ$}} &
\raisebox{-0.35ex}{\scalebox{1.75}{$\circ$}} &
\raisebox{-0.35ex}{\scalebox{1.75}{$\circ$}} &
{\scalebox{0.9}{\ding{108}}}  \\
\hline

{SecLLMHolmes}~\cite{secllmholm} &
\raisebox{-0.35ex}{\scalebox{1.75}{$\circ$}} &
\raisebox{-0.35ex}{\scalebox{1.75}{$\circ$}} &
\raisebox{-0.35ex}{\scalebox{1.75}{$\circ$}} & 
{\scalebox{0.9}{\ding{108}}} & 
\raisebox{-0.35ex}{\scalebox{1.75}{$\circ$}} &
{\scalebox{0.9}{\ding{108}}} &
{\scalebox{0.9}{\ding{108}}} &
\raisebox{-0.35ex}{\scalebox{1.75}{$\circ$}} &
\raisebox{-0.35ex}{\scalebox{1.75}{$\circ$}} &
\raisebox{-0.35ex}{\scalebox{1.75}{$\circ$}} &
\raisebox{-0.35ex}{\scalebox{1.75}{$\circ$}} & 
\raisebox{-0.35ex}{\scalebox{1.75}{$\circ$}} &
\raisebox{-0.35ex}{\scalebox{1.75}{$\circ$}} & 
\raisebox{-0.35ex}{\scalebox{1.75}{$\circ$}} &
{\scalebox{0.9}{\ding{108}}} &
{\scalebox{0.9}{\ding{108}}} \\
\hline

\multicolumn{17}{|c|}{\textbf{Automation Tools}} \\
\hline

{IRCopilot}~\cite{lin2025ircopilot} &
{\scalebox{0.9}{\ding{108}}} & 
\raisebox{-0.35ex}{\scalebox{1.75}{$\circ$}} &
\raisebox{-0.35ex}{\scalebox{1.75}{$\circ$}} &
\raisebox{-0.35ex}{\scalebox{1.75}{$\circ$}} & 
\raisebox{-0.35ex}{\scalebox{1.75}{$\circ$}} &
\raisebox{-0.35ex}{\scalebox{1.75}{$\circ$}} &
{\scalebox{0.9}{\ding{108}}} &
\raisebox{-0.35ex}{\scalebox{1.75}{$\circ$}} &
{\scalebox{0.9}{\ding{108}}} &
{\scalebox{0.9}{\ding{108}}} &
{\scalebox{0.9}{\ding{108}}} & 
\raisebox{-0.35ex}{\scalebox{1.75}{$\circ$}} &
\raisebox{-0.35ex}{\scalebox{1.75}{$\circ$}} & 
{\scalebox{0.9}{\ding{108}}} &
\raisebox{-0.35ex}{\scalebox{1.75}{$\circ$}} &
{\scalebox{0.9}{\ding{108}}} \\
\hline

{DeGPT}~\cite{hu2024degpt} &
\raisebox{-0.35ex}{\scalebox{1.75}{$\circ$}} &
\raisebox{-0.35ex}{\scalebox{1.75}{$\circ$}} &
\raisebox{-0.35ex}{\scalebox{1.75}{$\circ$}} & 
{\scalebox{0.9}{\ding{108}}} & 
\raisebox{-0.35ex}{\scalebox{1.75}{$\circ$}} &
\raisebox{-0.35ex}{\scalebox{1.75}{$\circ$}} &
{\scalebox{0.9}{\ding{108}}} &
\raisebox{-0.35ex}{\scalebox{1.75}{$\circ$}} &
\raisebox{-0.35ex}{\scalebox{1.75}{$\circ$}} &
\raisebox{-0.35ex}{\scalebox{1.75}{$\circ$}} &
{\scalebox{0.9}{\ding{108}}} & 
\raisebox{-0.35ex}{\scalebox{1.75}{$\circ$}} &
\raisebox{-0.35ex}{\scalebox{1.75}{$\circ$}} & 
\raisebox{-0.35ex}{\scalebox{1.75}{$\circ$}} &
{\scalebox{0.9}{\ding{108}}} &
{\scalebox{0.9}{\ding{108}}} \\
\hline

{PentestGPT}~\cite{deng2024pentestgpt} &
\raisebox{-0.35ex}{\scalebox{1.75}{$\circ$}} &
{\scalebox{0.9}{\ding{108}}} &
\raisebox{-0.35ex}{\scalebox{1.75}{$\circ$}} & 
\raisebox{-0.35ex}{\scalebox{1.75}{$\circ$}} & 
\raisebox{-0.35ex}{\scalebox{1.75}{$\circ$}} &
\raisebox{-0.35ex}{\scalebox{1.75}{$\circ$}} &
{\scalebox{0.9}{\ding{108}}} &
{\scalebox{0.9}{\ding{108}}} &
\raisebox{-0.35ex}{\scalebox{1.75}{$\circ$}} &
{\scalebox{0.9}{\ding{108}}} &
{\scalebox{0.9}{\ding{108}}} &
\raisebox{-0.35ex}{\scalebox{1.75}{$\circ$}} & 
\raisebox{-0.35ex}{\scalebox{1.75}{$\circ$}} &
{\scalebox{0.9}{\ding{108}}} &
\raisebox{-0.35ex}{\scalebox{1.75}{$\circ$}} &
{\scalebox{0.9}{\ding{108}}} \\
\hline
\end{tabular}
}
\caption{Comparison between State-of-the-art works.}
         \vspace{-0.3in}
\label{tab:comparison}
\end{table*}

\subhead{Comparison} 
Existing works on evaluating LLMs for security applications
can be broadly categorized based on two objectives: i) the development of LLM-based automation tools, and ii) the design of benchmarking frameworks for systematic evaluation. 
Benchmarking works are two types:
agentic (where an LLM agent interacts with an environment and performs actions)
and non-agentic (that involves static evaluation of knowledge and reasoning capabilities).
Automation tool development primarily focuses on maximizing performance on specific evaluation datasets, often through complex agent design or fine-tuning. In contrast, benchmarking frameworks aim to provide consistent and generalizable assessments across a range of use cases without prioritizing top scores. 
Thus, while LLM-based automation tools and benchmarking frameworks are complementary—each emphasizing either the tool or the data—they often share common components, such as LLM agents and datasets. Table \ref{tab:comparison} compares both categories
(where each comparison criteria is further explained below).

Our work focuses specifically on Security Incident Analysis tasks with multiple dataset design and agentic workflow features, with evaluation of both benign and attack artifacts, addressing a gap not covered by existing other security domain benchmarks.


\subhead{Comparison Criteria}
The first five columns compare works based on their specific application domains.
The following five columns compare dataset generation characteristics: \textit{Data-Refinement} indicates 
efforts such as paraphrasing, removing bias or crafting scenarios, to make tasks more real-life aligned. 
\textit{Increasing Task Difficulty} includes tasks with multiple difficulty levels;
\textit{Multiple Domain Coverage} checks if tasks span different security domains; 
\textit{Multi Steps} refers to tasks requiring multiple interactions; 
and \textit{Multiple Objectives} means analyzing multiple pieces of evidence. 
The next four columns compare based on the agentic workflow. The \textit{LLM agent} column indicates if a work develops an LLM-based system integrating LLM APIs. \textit{Environment Access} means if the LLM can interact with the environment (e.g., Kali). The \textit{Tool Access} column refers to the capability of the agents using tools to interact with the outer environment. The \textit{Context Optimize} column means to optimize the outputs provided to the agent for better decision making and reducing hallucination problems.
The next two columns indicate the type of target or sample being evaluated. \textit{Actual-Benign (Non-Vuln.)} refers to benign artifacts, benign code samples, or non-vulnerable machines. \textit{Actual-Malicious (Vuln.)} refers to malicious files, code, or vulnerable systems.

\section{Effect of Debiasing Strategies on Model Performance}
\label{sec:effect_debias}
Debiasing strategies can influence model performance in two ways: they can either help the LLM reason more effectively or make the task harder. Below, we present illustrative examples for both scenarios.

\subsection{When Debiasing Improves LLM Reasoning}

To demonstrate the impact of our question debiasing strategy (open-ended reformulation), we present an example using GPT-4o-mini on a network forensics task. The original (biased) question and its paraphrased (debiased) version are shown below, along with the corresponding model traces.

The task is to identify the IP address responsible for port-scanning activity in a packet-capture file. The original question assumes that port scanning has already occurred: \textit{``What is the IP responsible for conducting the port scan activity?''} In contrast, the debiased version removes this assumption and encourages the model to reason step by step: \textit{``Is there any evidence of port scanning in the network traffic? If so, what is the IP address responsible for the scanning activity?''}

\begin{center}
LLM Reasoning Without Question Debiasing
\end{center}
\vspace{-5pt}
\begin{lstlisting}[language=bash, basicstyle=\footnotesize\ttfamily, breaklines=true, frame=single]
# Without Debiasing of the questions:
Thought: To identify the source IP responsible for potential port scanning activity, 
I will analyze the provided capture.pcap file using tshark. This tool can help extract 
relevant information about the network traffic, including source IP addresses.
Action Input: tshark -r capture.pcap -q -z conv,ip
Output: 
**Source IP Addresses**:
   - The primary source IP involved in the communication is **10.251.96.5**.
   ...
   - The destination IP for the significant traffic is **10.251.96.4**.
   ...
Final Answer: The primary IP responsible for port scanning activity is "10.251.96.5."
\end{lstlisting}

\begin{center}
LLM Reasoning With Question Debiasing
\end{center}
\vspace{-5pt}
\begin{lstlisting}[language=bash, basicstyle=\footnotesize\ttfamily, breaklines=true]
# With Debiasing of the questions:
Thought: To determine if there is evidence of port scanning in the network traffic, 
I will analyze the provided capture.pcap file using tshark. Port scanning typically 
involves multiple connection attempts to various ports from a single IP address. 
I will look for patterns of SYN packets that indicate such activity.
Action Input: tshark -r capture.pcap -T fields -e ip.src -e ip.dst -e tcp.dstport | sort | uniq -c | sort -nr
Output:
**Source IP Addresses**:
   - The IP address responsible for conducting the port scan activity is **10.251.96.4**. 
     This IP has the highest number of connections (6365) to another internal IP, 
     **10.251.96.5**, primarily on port **80**.
   ...
   - The output shows a variety of connection counts for different ports, with the highest 
     counts indicating a focused scanning effort on specific services.
   ...
Final Answer: The primary IP responsible for port scanning activity is "10.251.96.4."
\end{lstlisting}

This example highlights how question debiasing improves the model's reasoning quality. In the original prompt, the question implicitly assumes that port scanning is present and directly requests the responsible IP. This presupposition biases the LLM toward selecting an IP address without fully verifying the underlying network behavior.

After debiasing, the paraphrased question removes the assumption and instead prompts the model to first determine whether scanning exists, and only then identify the scanning source. This encourages a more analytical reasoning process. As a result, the model chooses a more appropriate analysis method (per-port connection analysis) and correctly attributes the scanning activity to \texttt{10.251.96.4}, rather than incorrectly selecting \texttt{10.251.96.5} based on traffic volume.

\subsection{When Debiasing Makes the Task Harder}

To illustrate the opposite effect, we present an example using Claude-3.5-Sonnet on a memory forensics task. The goal is to identify the downloaded file by the attacker. The original question is: \textit{``What is the full URL of the PHP file that the attacker visited?''} The debiased version asks: \textit{``Can you identify if there is any evidence that the attacker visited a URL pointing to a file? If so, what is the full URL of the file that was accessed?''}

\begin{center}
LLM Reasoning Without Question Debiasing
\end{center}
\vspace{-5pt}
\begin{lstlisting}[language=bash, basicstyle=\footnotesize\ttfamily, breaklines=true]
# Without Debiasing of the questions:
Thought: To find the full URL of the PHP file visited by the attacker, 
I should examine strings in memory, particularly focusing on URLs and HTTP 
requests. I willl use the strings command combined with grep to search for PHP URLs.
Action Input: strings capture.mem | grep -i "\.php"
Output: 
**Source IP Addresses**:
   - The command output shows multiple PHP URLs, and among them, one notable URL appears to be:
`http://77.91.124.20/store/games/index.php`
   ...
Final Answer: The answer is: http://77.91.124.20/store/games/index.php"
\end{lstlisting}

\begin{center}
LLM Reasoning With Question Debiasing
\end{center}
\vspace{-5pt}

\begin{lstlisting}[language=bash, basicstyle=\footnotesize\ttfamily, breaklines=true]
# With Debiasing of the questions:
Thought: To find evidence of URLs accessed by the attacker, 
I will need to search through the memory dump for URLs. I can use 
the strings command combined with grep to search for common URL 
patterns, particularly focusing on file downloads.
Action Input: strings capture.mem | grep -i "http://" | grep -i "\.exe\|\.dll\|\.zip\|\.rar\|\.7z\|\.msi"
Output:
- System DLL files like httpprxc.dll, winhttp.dll, urlmon.dll
- HTTP system drivers (http.sys)
- CryptnetUrlCache directories for different applications and users
- URL monitoring and HTTP-related system components
However, the output does not reveal any specific URLs pointing to files 
that were accessed by the attacker. The results only show system files 
and cache locations rather than actual URL access evidence.
   ...
Final Answer: After trying multiple approaches to find evidence of URLs 
pointing to files (using strings search, filescan, and cmdline analysis), 
I have not found conclusive evidence of a specific URL pointing to a file 
that was accessed by the attacker. So, I am giving up on this question. 
\end{lstlisting}

In this example, debiasing the question actually makes the task harder for the LLM. The original question directly asks for the URL of the PHP file visited by the attacker. This gives the model a clear target and allows it to focus its search on PHP files in the memory dump, quickly identifying the relevant URL.

The debiased question, on the other hand, removes the assumption that a PHP file was visited and asks the model to first check for any evidence of file access. This broader framing increases uncertainty and makes the reasoning task more open-ended. As a result, the model spends effort scanning system files, caches, and various URL patterns, but fails to converge on the correct URL. 

\section{Ethical Considerations}
\label{appendix:ethics}

We decided to publish our SIA benchmark dataset, autonomous evaluation agent, and results publicly, upon acceptance. Specifically, we first discuss our compliance with the terms and conditions (T\&C) of the platforms from which we curated our data. Second, we identify both positive and negative outcomes that may arise from making the work publicly available, the risk mitigation steps taken, and how we weighed the benefits against the negative outcomes to arrive at this decision.

Our SIA dataset has been developed by carefully curating security incident-related challenges from three blue team platforms: BlueTeamLabsOnline \cite{BTLO}, CyberDefenders \cite{CyberDefenders}, and TryHackMe \cite{tryhackme}. Before data curation, we carefully reviewed their T\&C and complied with them by taking several steps, as detailed below. Additionally, we reviewed prior leading works, such as PentestGPT \cite{deng2024pentestgpt} and InterCode-CTF \cite{yang2023language}, who curated data from similar platforms as ours, and tailored their best practices to our released SIA dataset. PentestGPT and InterCode-CTF used challenges from HackTheBox \cite{hackthebox} and PicoCTF \cite{picoctf}, which abide by similar T\&C as ours. PentestGPT used multiple active labs from HackTheBox in their benchmark dataset and provided the names of the labs, allowing users instantiate the labs and run experiments. Since PentestGPT used active labs, all of them require login and some of them require paid subscriptions as well. The InterCode-CTF data file (.json) contains the CTF challenges (query and answer) and corresponding source URL referring to the original platform, and in some cases the URLs of the write-ups where they curated the solutions from.

Our problem set takes a similar approach. We provide the SIA challenges within our framework (e.g., scenario, questions, answers), but do not host files/artifacts directly and instead provide source URLs to access or download the files/artifacts from. Accessing the files/artifacts from the source platforms may require login, similar to the challenges from HackTheBox\cite{hackthebox} and picoCTF \cite{picoctf}, but do not require any subscription or payment. All SIA problems in our dataset are either retired or free; they have legit and multiple open-source write-ups \cite{Panagiotis, forensicskween, Medium}(under AGPL 3.0 license or open-source blogs) containing the scenario, questions, solving strategy, and answers. We provide URLs to all publicly available write-ups for our problem sets so users can verify the solutions. All these steps will facilitate the reproduction of the results without violating the T\&C of the platforms. That said, some files/artifacts involved may contain potentially harmful content. We strongly recommend executing the code and downloading artifacts only within isolated environments, such as a Kali Linux virtual machine, to ensure system safety.

For our Alert Triaging dataset, we developed it using publicly available files (PCAPs) from the CIC-IDS2017 and TII-SRC-23 dataset \cite{cicids2017, herzalla2023tii}. We carefully reviewed the terms of use associated with both CIC-IDS2017 and TII-SRC-23, which permit free usage for research and academic purposes. In accordance with these terms, we acknowledge the dataset, and cite the corresponding publications \cite{cicids2017, herzalla2023tii}.

By establishing an evaluation benchmark, our work highlights both the strengths and weaknesses of large language models (LLMs) in SIA tasks. It provides insights into where these models excel, where they struggle, and offers analyses of the reasons behind their failures—ultimately identifying areas that need further improvement to perform SIA tasks. While these findings are valuable, we acknowledge that some stakeholders may perceive them as a deterrent to rapid AI adoption, especially since the evaluation data clearly show that some models outperform others, and in certain contexts LLMs may not yet be mature enough. However, hasty adoption can backfire, resulting in financial losses, reputational harm, or even legal issues, particularly given the high-stakes nature of deploying AI in cyber defence. Therefore, our goal is to help the stakeholders make better-informed decisions by understanding the capabilities and limitations of LLMs in performing SIA tasks, thereby ensuring a more reliable return on investment. This knowledge also empowers them to critically assess AI tools, identify potential remedies or workarounds, and adopt them more responsibly. Interestingly, commercial companies like Anthropic have also recognized the value of public benchmarks, such as Cybench \cite{zhang2024cybench}, using them alongside internal evaluations and publishing the results in their reports. Having accessible means of independent validation fosters greater trust in AI models as well as the products built on top of them - leading to more robust and reliable use of AI in the long run.

We also acknowledge the potential risk of misuse associated with releasing our data, agent, and results. Malicious actors could adapt the agent for adversarial purpose. The results as well could inform them about the current capabilities of LLMs in solving SIA tasks, which knowledge could be exploited to deceive cyber defenders. However, given the existence of publicly available Capture The Flag (CTF) and penetration testing benchmarks and agents, we believe the risk of adapting our agent for adversarial purpose is low. Additionally, malicious actors may obtain similar technology through covert means. For these reasons, we believe accelerating cyber defense through advanced technologies like LLMs is critical. Our results track the progress of current LLMs on SIA tasks, identifies gaps, which would accelerate and foster the development of stronger AI agents for cyber defense.
In addition, as researchers, we prioritize transparency and reproducibility. By sharing our data, agent, and results, we aim to support the broader community in validating and building upon our work. Although due to the inherent variability in LLM outputs, full replication of our results may not always be possible \cite{muennighoff2025s1}.

After weighing the benefits such as advancing progress, informing stakeholders and policymakers, and enabling transparency and reproducibility over potential risks, we decided to publicly release our dataset, agent, results, and agent source code, upon acceptance of the paper.






%




\end{document}